\documentclass[12pt,a4paper]{article}
\usepackage{graphicx}
\usepackage{amssymb}
\usepackage{amsmath}
\usepackage{bm}
\usepackage{color}
\usepackage{theorem}
\usepackage{subfigure}

\usepackage[sort&compress,numbers, merge]{natbib}

\definecolor{Orange}{cmyk}{0,0.61,0.87,0}
\definecolor{JungleGreen}{cmyk}{0.99,0,0.52,0}
\definecolor{OliveGreen}{cmyk}{0.64,0,0.95,0.40}
\definecolor{Brown}{cmyk}{0,0.81,1,0.60}
\definecolor{RoyalBlue}{cmyk}{0.71,0.53,0,0.12}

\newcommand{\be}{\begin{equation}}
\newcommand{\ee}{\end{equation}}
\newcommand{\bea}{\begin{eqnarray}}
\newcommand{\eea}{\end{eqnarray}}
\newcommand{\zuv}{z_{\rm UV}}
\newcommand{\zir}{z_{\rm IR}}

\allowdisplaybreaks[1]

\newcommand{\1}{\mbox{1}\hspace{-0.25em}\mbox{l}}

\begin{document}

\begin{titlepage}
\begin{center}
\hfill {UMN--TH--4102/21}

\vspace{2.0cm}
{\Large\bf
Small Instantons  in Weakly-Gauged\\
\vspace{0.40cm}
Holographic Models}

\vspace{1.0cm}
{ \bf Tony Gherghetta$^{a}$ and
Alex Pomarol$^{b,c}$}

\vspace{0.5cm}
{\footnotesize\it
${}^a$School of Physics and Astronomy, University of Minnesota, Minneapolis, Minnesota 55455, USA\\
${}^b$IFAE and BIST, Universitat Aut\`onoma de Barcelona, 08193 Bellaterra, Barcelona\\
${}^c$Departament de F\'isica, Universitat Aut\`onoma de Barcelona, 08193 Bellaterra, Barcelona}

\vspace{0.5cm}
\abstract
Small instantons can play an important role in Yang-Mills theories whose gauge couplings are 
sizeable at small distances.  An interesting class of theories where this could occur is 
in weakly-gauged holographic models (dual to Yang-Mills theories interacting with strongly-coupled CFTs), 
since gauge couplings are indeed enhanced towards the UV boundary of the 5D AdS space.
However, contrary to expectations, we show that small instantons in these non-asymptotically-free models are highly suppressed and ineffective. This is due to the conservation of topological charge that forbids instantons to be localized near the UV boundary. Despite this fact we find non-trivial UV localized instanton-anti-instanton solutions of the Yang-Mills equations where the topological charges annihilate in the AdS bulk. These analytic solutions arise from a 5D conformal transformation of the uplifted 4D instanton. Our analysis therefore reveals unexpected nonperturbative configurations of Yang-Mills theories when they interact with strongly-coupled  CFTs.

\end{center}
\end{titlepage}
\setcounter{footnote}{0}


\section{Introduction}

Instantons play a prominent role in Yang-Mills theories, with their contribution to the path integral $\propto e^{-8\pi^2/g^2}$ becoming especially important for large gauge couplings. 
For example, in QCD, the Euclidean action of the BPST instanton~\cite{Belavin:1975fg}, after including  quantum corrections, is given by $S_E=8\pi^2/g^2(1/\rho)$ with a running gauge coupling $g(1/\rho)$ dependent on the instanton size $\rho$. Thus at large $\rho$, corresponding to IR energy scales where the QCD gauge coupling is large, nonperturbative effects due to ``large" instantons can dominate the path integral and lead to a variety of phenomena. 
  
This instanton dependence on the gauge coupling suggests that if the gauge coupling were to increase (or at least remain sizeable) at small distances (or UV energy scales) then small-size instantons could also dominate the path integral. This can be easily achieved by adding extra colored matter fields at some high-energy scale $m$, which changes the sign of the $\beta$-function, leading to a sizeable gauge coupling at small distances $\ll 1/m$. The matter fields can be either fermions or scalars. However, when the extra states are colored fermions, the axial anomaly causes the instanton contributions to explicitly depend on the fermion masses, $\propto (\rho\, m_F)^{N_F}$ (assuming a common mass $m_F$ and  number of fermions $N_F$), which again suppresses the small instanton contributions corresponding to $\rho\ll 1/m_F$! On the other hand, there is no extra suppression from the axial anomaly if the matter fields are colored scalars, and therefore small instanton contributions could be enhanced. 
However, in this case the scalar masses, $m_S$ must be tuned to be small,  $m_S\ll 1/\rho$,  in order to significantly  affect the running of the gauge coupling. Thus, it is not so straightforward to enhance the small instanton contributions of a Yang-Mills theory in a natural way.\footnote{
Alternatives that enhance the action $S_E$ from small instantons which is then used to increase the QCD axion mass are given in~\cite{Agrawal:2017ksf, Agrawal:2017evu,Csaki:2019vte,Gherghetta:2020keg}. In particular, small-size instantons are enhanced by  power-law corrections if  QCD propagates in a flat extra dimension at energies much larger than the electroweak scale~\cite{Gherghetta:2020keg}.}

A third possibility to enhance the small instanton contributions is to consider QCD propagating in five-dimensional (5D) Euclidean AdS.  In this theory, gluons  on the UV boundary of the AdS space
receive a well-known (classical) logarithmic contribution~\cite{Pomarol:2000hp, Arkani-Hamed:2000ijo, Goldberger:2002cz} that is purely due to the AdS geometry and does not rely on introducing extra colored states. Importantly, this contribution enhances the gauge coupling  of the gluons on the UV boundary at small distances. 

This logarithmic effect has a 4D interpretation based on  the AdS/CFT correspondence~\cite{Maldacena:1997re}. The 5D AdS QCD theory is dual to a 4D strongly-coupled CFT with a global SU(3)$_c$ symmetry that contains a large number $N$ of constituents charged under this SU(3)$_c$. By weakly gauging SU(3)$_c$ (identified as QCD color), this CFT charged matter would then be responsible for causing the QCD coupling to grow at small distances. However, we should emphasize that this is only a holographic interpretation of the 5D logarithmic running and there are no charged scalar or fermion constituents in the 5D model. Thus, the inherent drawbacks of extra colored matter fields, discussed previously, could in principle be avoided.

Furthermore, by the AdS/CFT correspondence, we expect 5D holographic models to have an instanton action similar to that found in 4D theories, $S_E=8\pi^2/g^2(1/\rho)$, where $g(1/\rho)$ is identified with the running coupling of the gluons on the UV boundary. This expected form of the action can be directly interpreted in the 5D theory by using the fact that the AdS/CFT dictionary relates the renormalization scale $\mu$ of the running coupling $g(\mu)$ to the inverse of the 5th coordinate $z$, i.e. $\mu\sim 1/z$. Given that the running coupling in the 4D instanton action is renormalized at $\mu=1/\rho$, therefore suggests that in the 5D AdS-Yang-Mills theory there must be an instanton configuration localized on the UV boundary that extends only a finite distance $z\sim \rho$ into the 5D bulk.

\begin{figure}[t]
\begin{center}
\includegraphics[width=.53\textwidth]{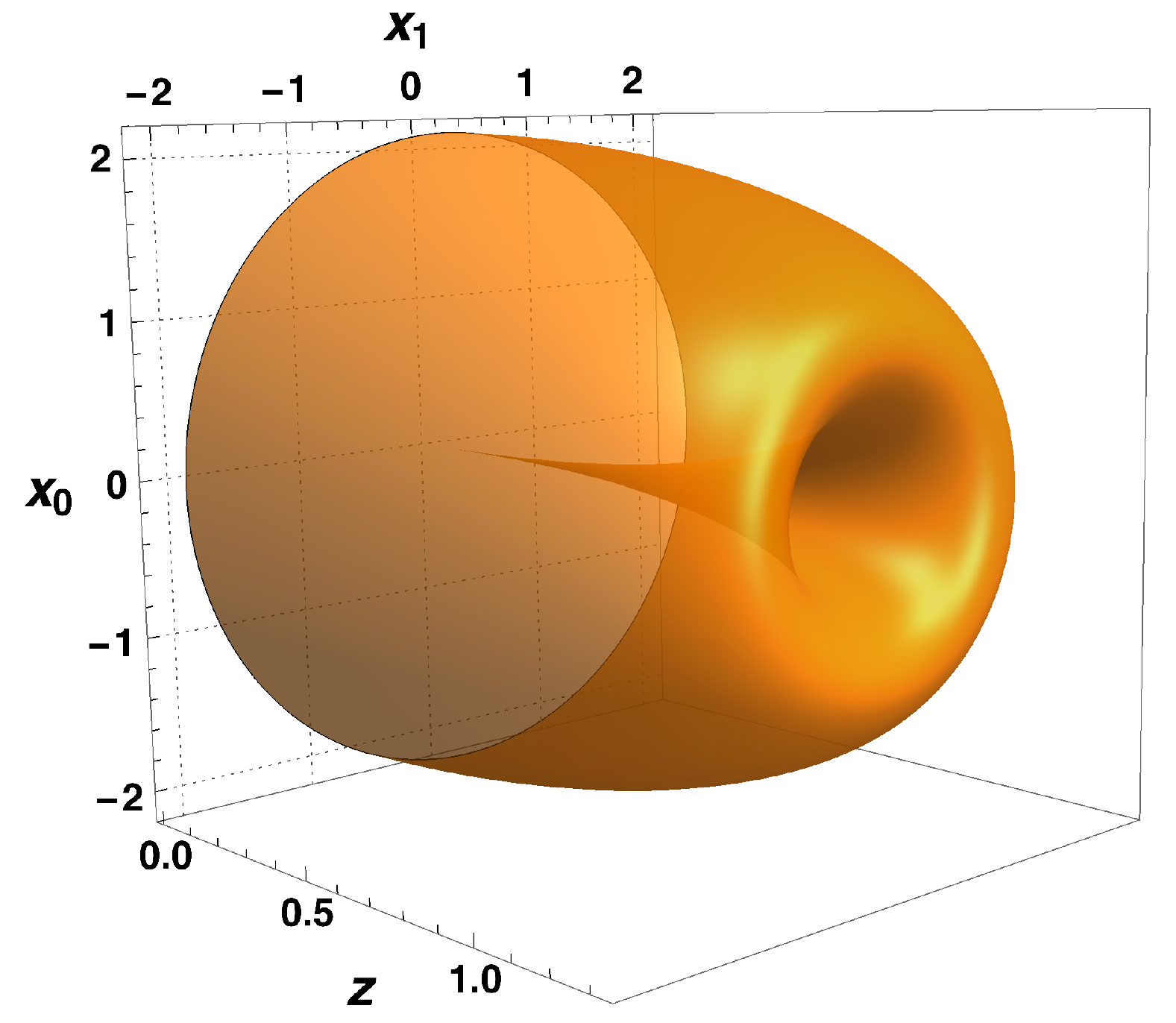}
\caption{A 2D contour  of the 5D Lagrangian density \eqref{eq:energy4DIinv} (multiplied by the volume factor $2\pi^2 x^3$ and for $\rho=1$) evaluated at 10, depicting an instanton and anti-instanton both located at $x_\mu=0$ (where two Euclidean coordinates $x_{2,3}$ are not shown). As $z$ increases, 
the instanton size remains fixed while the anti-instanton size grows until it annihilates the instanton at $z\sim \rho$.}
\label{fig:3dplot4diai}
\end{center}
\end{figure}

To look for such a solution we consider an SU(2) Yang-Mills theory in Euclidean AdS$_5$ whose limit on the AdS boundary is a 4D instanton. However, it turns out that such a solution is, in fact, {\it not} possible because topological charge conservation implies that the 4D instanton cannot ``disappear" into the AdS 
bulk (as can also be directly verified by examining the 5D AdS Yang-Mills equations).  
Indeed the only instanton solution to the 5D AdS Yang-Mills equations that we obtain is from uplifting the 4D instanton to the AdS$_5$ bulk as a ``cylindrical" configuration with no dependence on the 5th coordinate. The corresponding topological charge, defined on 4D slices at any given $z$,  is then always one for this solution. Since this uplifted instanton solution is not localized, we obtain $S_E=8\pi^2/g^2(\zir^{-1})$, where $\zir\gg \rho$ is an IR cutoff  (identified with the confinement scale of the CFT). 
This means that despite the fact that the gauge coupling increases towards the UV boundary, small instantons are highly suppressed and ineffective. 

Alternatively, this suppression can be understood from a supersymmetric argument. Interestingly, the ``cylindrical" instanton configuration is also a solution of the supersymmetrized 5D model, and therefore by the AdS/CFT dictionary implies that the dual 4D theory necessarily contains colored {\it fermion} constituents. The axial anomaly of these colored fermions then leads to the usual chiral suppression (but saturated at $\zir^{-1}$). At large $N$, where the effects from the colored CFT fermions dominate, this suppression can be simply interpreted in the 5D theory as an extra running of the gauge coupling from $1/\rho$ to $\zir^{-1}$ to give the corresponding action $S_E=8\pi^2/g^2(\zir^{-1})$. Thus, the absence of an expected localized instanton solution results from an accidental supersymmetry of the bulk at the classical level.

These arguments therefore suggest that AdS$_5$ Yang-Mills theory only admits UV localized configurations with zero topological charge. Indeed, we discover new instanton-anti-instanton configurations of the 5D AdS Yang-Mills theory which extend only a finite distance into the 5D bulk. In particular, we find a solution that for $z\ll \rho$, consists of an anti-instanton of size $z^2/\rho$ located inside an instanton of size $\rho$. As $z$ increases, the anti-instanton grows in size until it ``pinches off" the instanton at a distance in the bulk $z\sim \rho$. A two-dimensional contour of the Lagrangian density 
of this solution is shown in Figure~\ref{fig:3dplot4diai}. This solution is analytically obtained by taking advantage of the SO(5,1) isometry group of Euclidean AdS$_5$, corresponding to the conformal symmetry of the dual 4D CFT. In particular, the instanton-anti-instanton solution is simply generated by performing a discrete 5D conformal transformation (or ``inversion") on the uplifted 4D instanton solution.

This localized zero topological charge solution can also be understood using the accidental supersymmetry in the bulk. However, contrary to the ``cylinder" instanton, the instanton-anti-instanton configuration does not preserve the 5D supersymmetry and consequently there are no fermion zero modes in the dual 4D theory (that would lead to extra chiral suppression from the axial anomaly). Thus, the running gauge coupling only receives contributions from the fermion nonzero modes, whose running ends at $1/\rho$ and not $\zir^{-1}$.

Our solution-generating technique can be extended by performing other 5D conformal transformations on known 4D instanton solutions. For instance, performing a 5D special conformal transformation on the uplifted 4D instanton solution gives rise to other instanton-anti-instanton solutions. In fact by performing multiple, independent, special conformal transformations on the 4D instanton or multi-instanton solutions~\cite{Actor:1979in}, an instanton-anti-instanton solution ``gas" can be generated. In all cases the topological charge is preserved in the AdS bulk. Similarly, we find generalizations of known 4D meron solutions~\cite{deAlfaro:1976qet,Callan:1977gz, deAlfaro:1977zs} to 5D configurations.
Curiously, we also discover a new, simple analytical solution to the 5D AdS Yang-Mills equations
that is {\it not} obtained from any 5D conformal transformation of a known 4D solution.  
It is a  5D meron configuration, that for any given $z$ has a topological charge $\frac{1}{2}$ and size proportional to $z$. Unlike the known 4D meron which is singular, the 5D meron is regular. Again, applying 5D special conformal transformations on the new 5D meron solution further generates other meron-antimeron configurations.

\section{Yang-Mills theory interacting with a strongly-coupled CFT}
\label{sec:holoYM}

We are interested in studying Euclidean topological configurations of Yang-Mills theories interacting with a strongly-coupled CFT. For simplicity, we consider an SU(2)  theory with gauge bosons $A_\mu^a (a=1,2,3)$ whose  4D Lagrangian is given by
\begin{equation}
  {\cal L}_4 = \frac{1}{4 g_4^2} F_{\mu\nu}^a F^{a\mu\nu}+A_\mu^a J_{\rm CFT}^{a\mu} +{\cal L}_{\rm CFT}~,
\label{eq:4dL}
\end{equation}
where $g_4$ is the Yang-Mills coupling, $J_{\rm CFT}^{a\mu}$ is the SU(2) CFT current and ${\cal L}_{\rm CFT}$ is the CFT Lagrangian. 

We will study the above Yang-Mills theory in the limit where it admits a 5D holographic description. In this limit, the CFT is described by gravity in AdS$_5$ with spacetime coordinates $x_M=(x_\mu,z)$ $(M=0,1,2,3,5)$ and metric
\begin{equation}
    ds^2=g_{MN} d x^M dx^N =\frac{L^2}{z^2} \left( \delta_{\mu\nu} d x^\mu dx^\nu + d z^2 \right)\equiv w^2(z)\left( \delta_{\mu\nu} d x^\mu dx^\nu + d z^2 \right)~,
    \label{eq:5dmetric}
\end{equation}
where $L$ is the AdS$_5$ curvature length, $\delta_{\mu\nu} = {\rm diag}(1,1,1,1)$ is the 4D Euclidean metric and the warp factor $w(z)=L/z$. Following the usual holographic procedure, the AdS$_5$ action is regularized by truncating the space with a boundary at $z=\zuv$ and taking the limit  $\zuv\to 0$
after the proper counterterms have been introduced on this UV boundary to cancel  UV divergences.
Also for convenience, we introduce an IR boundary at $z=\zir\gg \zuv$. This corresponds to an IR mass gap  in the CFT  at the scale $\sim z_{\rm IR}^{-1}$ that also helps to regularize IR divergences.

In the 5D holographic model the  SU(2) global symmetry of the CFT corresponds to introducing an SU(2) local symmetry in the bulk with gauge bosons $A_M^a$ $(a=1,2,3)$. The 5D Euclidean action is then given by
\begin{equation}
S_5= \int d^4 x \int_{\zuv}^{\zir} dz \sqrt{g}~\frac{1}{4 g_5^2} F_{MN}^a F^{aMN}~,
\label{eq:5DLag}
\end{equation}
where $F_{MN}^a=\partial_M A_N^{a}-\partial_N A_M^{a}+\varepsilon^{abc} A_M^b A_N^c$  is the field-strength tensor with $\varepsilon^{abc}$ the Levi-Civita symbol, and $g_5$
is the dimensionful 5D SU(2) gauge coupling.
The 4D Yang-Mills gauge bosons $A^a_\mu$, which are introduced in \eqref{eq:4dL}
as external  fields weakly coupled to the CFT, correspond in the 5D theory to the bulk gauge field evaluated on the UV boundary $A^a_\mu \equiv A^a_\mu(x,\zuv)$.  On this  boundary we can add the action 
\begin{equation}
	S_{\rm UV}= \int d^4 x~\frac{1}{4 g_4^2} F_{\mu\nu}^a F^{a\mu\nu}\Bigg|_{\zuv}~,
\label{eq:UVLag}
\end{equation}
where the $\mu,\nu$ indices are contracted with the 4D Euclidean metric $\delta_{\mu\nu}$
and  the boundary gauge coupling has  been identified with $g_4$ in \eqref{eq:4dL}. As will be shown, the boundary kinetic term  \eqref{eq:UVLag} acts as a counterterm to regularize the divergent 5D contributions to $A^a_\mu$ in the limit $\zuv\to0$. 

The bulk equation of motion can be obtained by  extremizing  \eqref{eq:5DLag}, $\delta S_5/\delta A^a_M=0$ which gives
\begin{equation}
        \partial_{M} \left( w(z) F^{MNa}\right) + w(z)\varepsilon^{abc} A_M^b F^{MN c} +{\cal B}_\mu^a \delta_{\mu}^N=0~,
       \label{eq:5Deom}
\end{equation}
where $\delta_\mu^N$ is the Kronecker delta and the boundary term is
\begin{equation}
       {\cal B}_\mu^a= w(z) F_{\mu 5}^a  \Big|_{\zuv}^{\zir}~.
        \label{eq:boundaction}
\end{equation}
The last term in \eqref{eq:5Deom},  together with the contribution from \eqref{eq:UVLag}, leads to the 
UV boundary condition
\begin{equation}
      \frac{1}{g_4^2}\left( \partial_{\mu}  F^{\mu\nu a} + \varepsilon^{abc} A_\mu^b F^{\mu\nu c}\right)-\frac{1}{g_5^2}w(z) F_{\nu 5}^a \Big|_{\zuv}=0~.
      \label{eq:4Deombdy}
\end{equation}
This corresponds to the equation of motion of a 4D Yang-Mills field coupled to a current that can be identified with the CFT current in \eqref{eq:4dL}:  $J_{\rm CFT}^{a\mu}= -\frac{1}{g_5^2}w(z) F_{\mu 5}^a|_{\zuv}$. The IR boundary condition is obtained from (6) and given by $F_{\mu5}^a|_{\zir} =0$.

We will be working in the limit in which the 5D coupling $g_5\to 0$ and 
gravitational interactions are neglected. 
This limit is well-defined and corresponds in the dual 4D theory to a large $N$ expansion with
the 5D gravitational coupling scaling as 
$G_{\rm N}\sim 1/N^2$ while $g_5^2\sim 1/N$. 

Using the above holographic model we can calculate the CFT contribution to the 4D Yang-Mills gauge coupling, as this will be useful later. This is done by integrating out the 5D gauge boson as a function of the boundary field $A_\mu^a$ at the quadratic level to give the UV-boundary term~\cite{Pomarol:2000hp}
\begin{equation}
	\int d^4 x  \frac{1}{4}\left(\frac{L}{g_5^2}\log\left(\frac{\zir}{\zuv}\right)\right)F_{\mu\nu}^a F^{a\mu\nu}\Bigg|_{\zuv}~.
\label{eq:loop}
\end{equation}
This term  diverges as $\zuv\to 0$.  However, this divergence can be cancelled by the counterterm  \eqref{eq:UVLag}. Indeed, by defining the ``running" gauge coupling as
\begin{equation}
      \frac{1}{g_4^2(\mu)} = \frac{1}{g_4^2} -\frac{L}{g_5^2}\log(\mu \zuv) ~,
      \label{eq:4drenorm}
\end{equation}
the bulk contribution  \eqref{eq:loop}  corresponds to the replacement $g_4^2\to g_4^2(\mu=z_{\rm IR}^{-1})$ in $\eqref{eq:UVLag}$. This is the expected result of an SU(2) Yang-Mills theory with $\beta$-function coefficient, $b_{\rm CFT}\equiv 8\pi^2\frac{L}{g_5^2}\sim N\gg 1$ due to the CFT\,\footnote{The $\beta$-function coefficient $b$ is defined from the renormalization group equation: $\frac{d}{d\log\mu} \left(\frac{8\pi^2}{g_4^2}\right) \equiv -b$. For SU(2) QCD this definition corresponds to $b_{QCD}=-22/3+2/3 n_f$, where $n_f$ is the number of Weyl fermions while the CFT contribution is $b_{\rm CFT} =8\pi^2 L/g_5^2$.}, that changes the Yang-Mills gauge coupling down to energies $\mu=z_{\rm IR}^{-1}$. 
Note that the running in \eqref{eq:4drenorm} only includes the CFT matter contributions, while the 4D gluon contributions,  which are sub-dominant in the large $N$ expansion, are neglected.

\section{AdS$_5$ Yang-Mills Euclidean configurations}
\subsection{Solution ansatz}
To obtain solutions to \eqref{eq:5Deom} we assume the following ansatz:
\begin{equation}
       A_\mu^a(x,z) = 2\eta_{\mu\nu}^a \frac{x_\nu}{x^2} f(x,z)~,\quad A_5^a(x,z)=0~,
 \label{eq:gansatz1}
\end{equation}
where $\eta_{\mu\nu}^a$ is the `t Hooft symbol~\cite{tHooft:1976snw}, $x^2=x_0^2+x_1^2+x_2^2+x_3^2$ and $f(x,z)$ is an arbitrary function. 
The matrix gauge potential corresponding to the ansatz \eqref{eq:gansatz1} can be written in terms of a gauge parameter 
$\Omega$ as
\begin{equation}
     A_\mu(x,z) =f(x,z) \left[-i (\partial_\mu \Omega) \Omega^{-1}\right]~,
     \label{eq:Agauge}
\end{equation}
where 
\begin{equation}
         \Omega =\frac{1}{\sqrt{x^2}} \left( {\vec x} \cdot {\vec\sigma} \mp i x_0\1 \right)~,
         \label{eq:omegadef}
\end{equation}
with $\Omega^{-1} = \Omega^\dagger$ and $\sigma_i$ ($i=1,2,3$) are the Pauli matrices. Note that the matrix gauge potential can be gauge transformed as 
\begin{equation}
         A_\mu' = U^\dagger A_\mu U -i (\partial_\mu U^\dagger) U~,
         \label{eq:gaugetransform}
\end{equation}
where $U=U(x)$ is a gauge transformation. 

Substituting the ansatz \eqref{eq:gansatz1} into the equation of motion \eqref{eq:5Deom} gives a second-order nonlinear partial differential equation for $f(x,z)$:
\begin{equation}
   f^{(2,0)} +\frac{1}{x}f^{(1,0)}+f^{(0,2)}-\frac{1}{z}f^{(0,1)} -\frac{4}{x^2} f(f-1)(2f-1)=0~,
   \label{eq:feom}
\end{equation}
where the first (second) index in the superscript denotes differentiation with respect to $x\ (z)$.
Note that if $f$ is a solution, then $1-f$ is also a solution.
With the ansatz \eqref{eq:gansatz1} we obtain the 5D Lagrangian density 
\begin{equation}
         \frac{1}{4} F_{MN}^aF^{aMN}= \frac{6}{x^2}\left((f^{(0,1)})^2 + (f^{(1,0)})^2   +\frac{4}{x^2} f^2(f-1)^2\right)~.
\end{equation}
At a fixed, nonzero $z$, corresponding to a 4D slice of AdS$_5$, we can also define a topological charge density by
\begin{equation}
    D= \frac{1}{4} F_{\mu\nu}^a {\widetilde F}^{a\mu\nu}= \frac{24}{x^3} f^{(1,0)} f(1-f)=\frac{4}{x^3}\frac{\partial}{\partial x} \left( 3f^2-2f^3\right)~,
    \label{eq:topdensitydefn}
\end{equation}
where ${\widetilde F}_{\mu\nu}^a =\frac{1}{2} \varepsilon_{\mu\nu\rho\sigma} F^{\rho\sigma a}$ with $\varepsilon_{\mu\nu\rho\sigma}$ the Levi-Civita symbol.
The topological charge $q$ on the 4D slice is then given by 
\begin{equation}
         q= \frac{1}{8\pi^2}\int d^4 x \,D~.
         \label{topologcharge}
\end{equation}
The solution $1-f$ has an opposite topological charge to $f$. For a 4D spherically symmetric solution, the topological charge
\eqref{topologcharge} can be written as
\begin{equation}
         q= (3f^2-2f^3)\Big|_0^\infty~.
         \label{topcharge}
\end{equation}
In particular, configurations with $q=1$  have $f$ changing from $f(0,z)=0$ to $f(\infty,z)=1$.

We next present solutions to the AdS$_5$ Yang-Mills equation of motion \eqref{eq:feom} by first considering the trivial case of $f(x,z)\equiv f(x)$, followed by nontrivial $z$-dependent solutions.

\begin{figure}[t]
\begin{center}
\includegraphics[width=.52\textwidth]{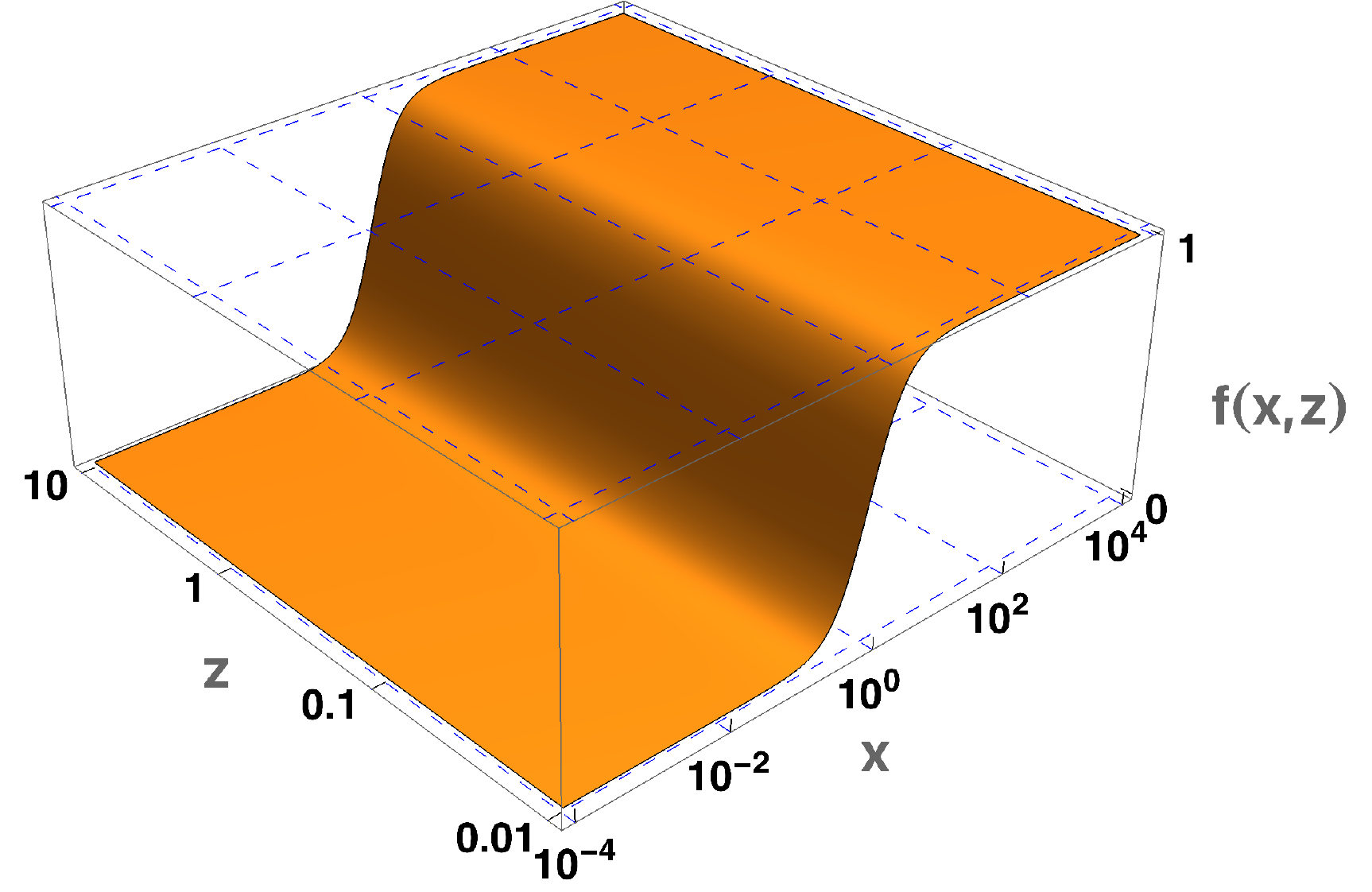}
\includegraphics[width=.47\textwidth]{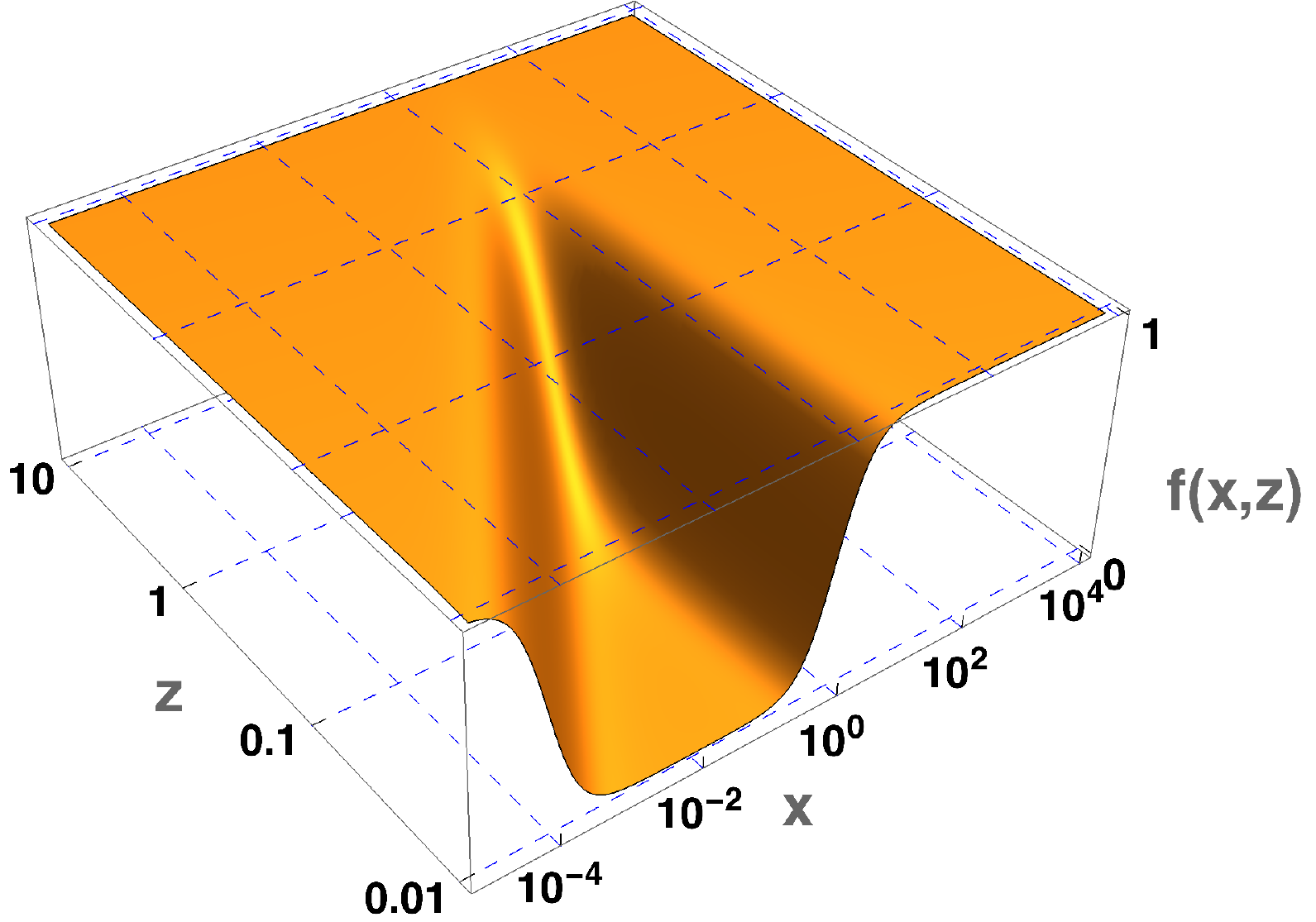}
\caption{Left: The uplifted 4D instanton  solution \eqref{eq:4dinstanton} for $\rho=1$, depicting the $f=0$ asymptotic region rising to $f=1$ at $x\sim \rho$, independent of $z$.
Right: The instanton-anti-instanton solution \eqref{eq:4dinstantonnew2} for $\rho=1$, depicting the merging of the asymptotic $f=1$ regions at $x\sim z\sim \rho$ from the anti-instanton and instanton. 
The $f=1$ ``plateaus" enclose a ``basin" corresponding to $f=0$.}
\label{fig:f5Dinv}
\end{center}
\end{figure}

\subsection{Uplifted 4D Instanton}
\label{sec:uplifted4DI}
A trivial 5D solution of \eqref{eq:feom} can be obtained by assuming that $f(x,z)$ is $z$ independent. 
This leads to the solution 
\begin{equation}
      f(x,z) = \frac{x^2}{x^2+\rho^2}~,
      \label{eq:4dinstanton}
\end{equation}
which is just the usual 4D instanton solution with size $\rho$ and topological charge $q=1$, uplifted to the fifth dimension. The solution \eqref{eq:4dinstanton} is plotted in Figure~\ref{fig:f5Dinv}, while the topological charge density, $D=48\rho^4/(x^2+\rho^2)^4$ is shown in Figure~\ref{fig:4dI5dinv}. 
The Lagrangian density is
\begin{equation}
     \frac{1}{4} F_{MN}^aF^{aMN} =  \frac{48\rho^4}{(x^2+\rho^2)^4}~,
     \label{eq:4DIenergy}
\end{equation}
which leads to the 5D action 
\begin{equation}
      S_5=8\pi^2 \frac{L}{g_5^2} \log\left(\frac{\zir}{\zuv}\right)~.
      \label{eq:5dactionlog}
\end{equation}
Similarly, an anti-instanton solution with topological charge $q=-1$ arises from the replacement $f \rightarrow 1-f$, and also leads to \eqref{eq:5dactionlog}. It is important to remark that for small but finite $g_5$, the cutoff of the 5D model, $\Lambda_5$ (where loop corrections become of order one) scales as 
$\Lambda_5\sim 1/g_5^2$ and then \eqref{eq:5dactionlog} is expected  to  be sensitive to the cutoff for $z\gtrsim \rho /g_5^2L$. Here we are only  considering the formal limit $g_5\to 0$ such that $\Lambda_5\to\infty$.

The 5D action \eqref{eq:5dactionlog} diverges as $\zuv\to 0$, but as outlined in Section~\ref{sec:holoYM}, this divergence can be  cancelled by the UV boundary term \eqref{eq:UVLag}   evaluated on the instanton solution \eqref{eq:4dinstanton} and using \eqref{eq:4drenorm}. Combining this with \eqref{eq:5dactionlog}, we obtain the finite action
\begin{equation}
      S_5+S_{\rm UV}=\frac{8\pi^2}{g^2_4(z_{\rm IR}^{-1})}~,
      \label{eq:5dtotal}
\end{equation}
where the Yang-Mills gauge coupling is now evaluated at the scale $z_{\rm IR}^{-1}$.\footnote{Recall from \eqref{eq:4drenorm} that only the CFT matter contribution to the gauge coupling running is considered, since this dominates at large $N$.}
This result is unexpected. As explained in the Introduction, when matter fields are added to a Yang-Mills theory, instantons receive loop corrections that can be captured by the replacement $g_4\to g_4(1/\rho)$~\cite{Vainshtein:1981wh}. However, the gauge coupling in \eqref{eq:5dtotal} does not exhibit this behaviour and instead the gauge coupling dependence gives a much larger action for small instantons with $\rho\ll z_{\rm IR}$, since $g_4(z_{\rm IR}^{-1})\ll g_4(1/\rho)$ according to \eqref{eq:4drenorm}.
Therefore, we find that small instantons in weakly-gauged holographic models are much more suppressed than expected. 

The reason for the larger than expected action \eqref{eq:5dtotal} boils down to the fact that the 5D instanton solution extends from $\zuv$ to $\zir$ (see Figure~\ref{fig:f5Dinv}) without any suppression. If the 5D instanton were localized towards the UV boundary, and extended only a distance $\rho$ in the bulk, the corresponding action would be of order $8\pi^2/g^2_4(1/\rho)$. However, we find that a UV localized instanton solution is not possible.  This is because such a solution would require the topological charge \eqref{topcharge} to change from one at $\zuv$, to zero at some $z>\zuv$, which is not consistent with topological charge conservation.\footnote{This can also be understood from the equation of motion \eqref{eq:feom}. A UV localized 5D solution would require that the gauge field configuration becomes pure gauge in the bulk (either $f\to 1$ or $0,~\forall x$) at some $z>\zuv$. But starting on the UV boundary with the 4D instanton solution \eqref{eq:4dinstanton}, $f(x,z)$ interpolates between $f=0$ at $x=0$ to $f=1$ at $x\rightarrow \infty$. Therefore, a UV localized instanton would require  $f$ to abruptly change at some finite $z>\zuv$ from $f=0\to 1$ at $x=0$ (or from $f=1\to 0$ at $x\rightarrow \infty$). 
Assuming $f$ is a smooth function, this behaviour is not possible as can be seen by, for example, taking the limit $x\rightarrow 0$ in \eqref{eq:feom}. In this limit, the nonlinear and $x$-derivative terms dominate over the $z$ derivative terms (assuming $f\simeq x^2/(x^2+\rho^2)$ corresponding to a 4D instanton), and therefore near $x=0$ the solution $f$ remains approximately constant as $z$ increases. This implies that $f$ (near $x=0$) cannot  change from $f=0\to 1$ along the $z$ direction. A similar argument holds, if the change $f=1\to 0$ were to occur in the limit $x\to \infty$. A loophole to this argument could be to allow for singular configurations. In this case a negative charged configuration emerging from the bulk could compensate the positive charge of the UV boundary instanton.}

Supersymmetry provides an alternative explanation of why the expected action $S_E=8\pi^2/g^2_4(1/\rho)$ was not obtained in the 5D holographic model. This follows by noting that the instanton configurations of our 5D model \eqref{eq:5DLag} are also solutions of the supersymmetric version~\cite{Gherghetta:2000qt}, since the supersymmetric partners (gluinos and adjoint scalars) can be consistently put to zero in the instanton configuration. This implies that these configurations preserve supersymmetry at the classical level, and, by the AdS/CFT correspondence, must be compatible with instanton solutions of  the 4D dual supersymmetric CFT at the quantum level. These theories necessarily contain massless colored fermions, $\psi$ that require a 't Hooft fermion vertex to be included in the partition function \cite{Shifman:1979uw}. Furthermore, these CFT fermions contribute instanton zero modes to the partition function but importantly the nonzero mode contributions cancel~\cite{Shifman:1999mv,Poppitz:2002ac}. This modifies the instanton measure to give~\cite{tHooft:1976snw}
\be
{\langle{\rm det}(\bar\psi_L\psi_R)\rangle}{\rho^{b_{\rm CFT}}} e^{-\frac{8\pi^2}{g^2_4(1/\rho)}}
\sim \left(\frac{\rho}{\zir}\right)^{b_{\rm CFT} } e^{-\frac{8\pi^2}{g^2_4(1/\rho)}}=
e^{-8\pi^2\left[\frac{1}{g^2_4(1/\rho)}-\frac{b_{\rm CFT}}{8\pi^2}\log(\rho/\zir)\right]}~,
\label{pi}
\ee
where $b_{\rm CFT} \propto N$ and $1/\zir$ is identified with the condensate scale of the CFT.  The negative of the exponent of \eqref{pi} then matches the result \eqref{eq:5dtotal} since $b_{\rm CFT}= 8\pi^2 L/g_5^2\sim N$ and gives the running from $1/\rho$ to $z_{\rm IR}^{-1}$ (again this assumes that the CFT matter dominates the running of the gauge coupling).  In other words, a 5D localized instanton with action $8\pi^2/g^2_4(1/\rho)$ cannot be a solution in a 4D supersymmetric CFT at the quantum level.

In summary, we see that (nonsingular) UV localized solutions are only possible if the topological charge is already zero on the UV boundary. This is possible for instanton-anti-instanton solutions as we discuss in the next section.

\subsection{5D Localized Instanton-Anti-Instanton solutions}

New solutions to the 5D AdS Yang-Mills equations can be generated by using the fact that the SU(2) gauge theory \eqref{eq:5DLag} and the AdS$_5$ metric are invariant under conformal transformations. This includes both discrete and continuous transformations which are given in Appendix~\ref{sec:App5D}.
By transforming \eqref{eq:4dinstanton}, new solutions to the 5D AdS Yang-Mills equations will be generated.

\subsubsection{An instanton-anti-instanton by 5D inversion}
\label{sec:5Diai}

Consider the uplifted 4D anti-instanton solution, $f(x,z)=\rho^2/(x^2+\rho^2)$. Under a discrete 5D inversion \eqref{eq:invtrans} with $\delta=\rho$, this solution transforms to
\begin{equation}
      f(x,z) =~\frac{(x^2+z^2)^2}{x^2\rho^2+(x^2+z^2)^2}~,
      \label{eq:4dinstantonnew2}
\end{equation}
where the matrix gauge parameter part contained in the square brackets of \eqref{eq:Agauge} remains invariant using \eqref{eq:Atransform}.
This is also a solution of \eqref{eq:feom}, and when $z=0$ we recover a 4D instanton solution.\footnote{This is expected because a 4D coordinate inversion transforms an anti-instanton into an instanton and vice-versa.} 
The solution \eqref{eq:4dinstantonnew2} is shown in Figure~\ref{fig:f5Dinv}.
Note that performing a 5D inversion on the uplifted 4D instanton solution would instead give rise to  the solution $1-f$ (with $f$ given in \eqref{eq:4dinstantonnew2}), which becomes an anti-instanton at $z=0$.

\begin{figure}[t]
\begin{center}

\includegraphics[width=.49\textwidth]{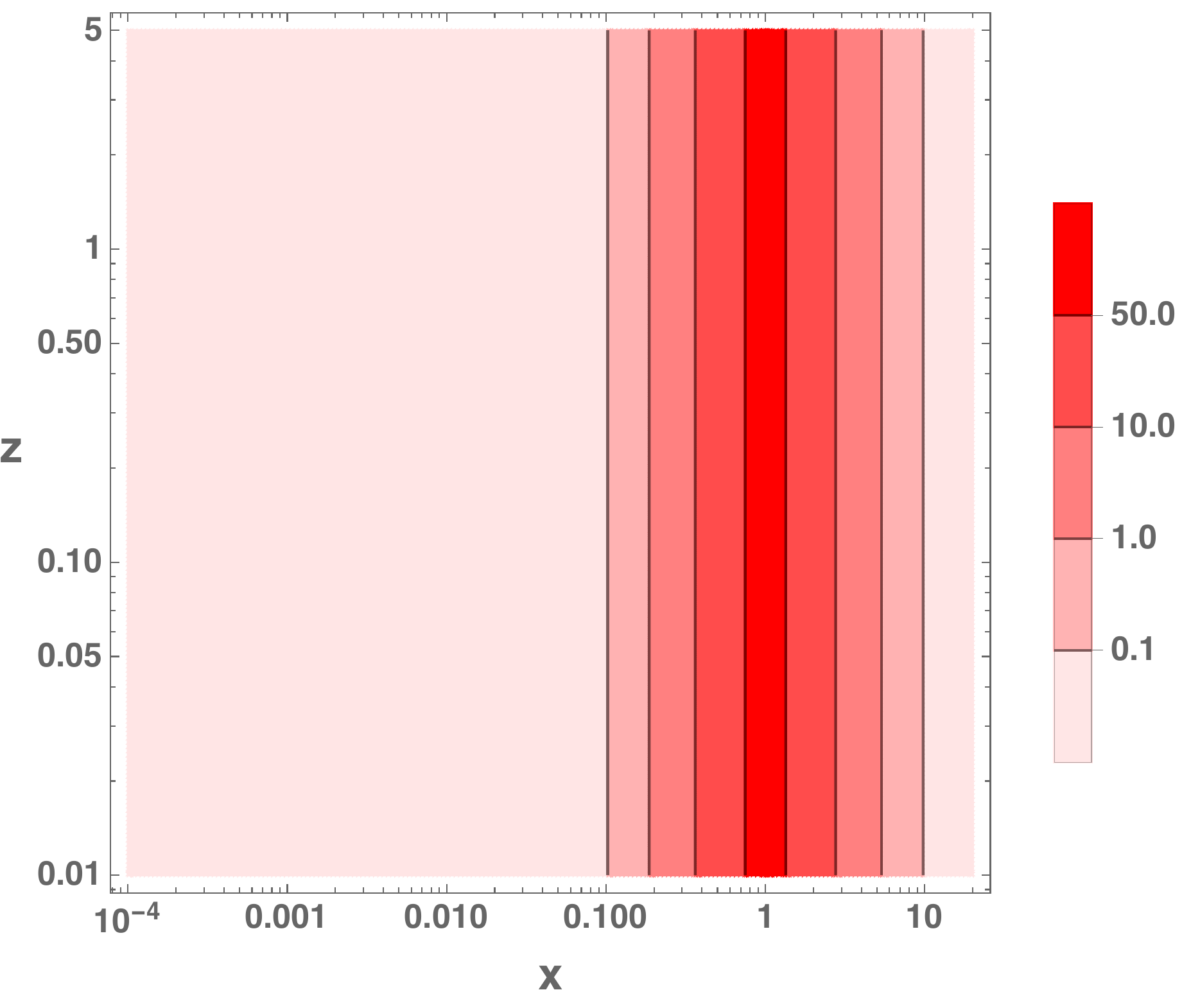}
\includegraphics[width=.49\textwidth]{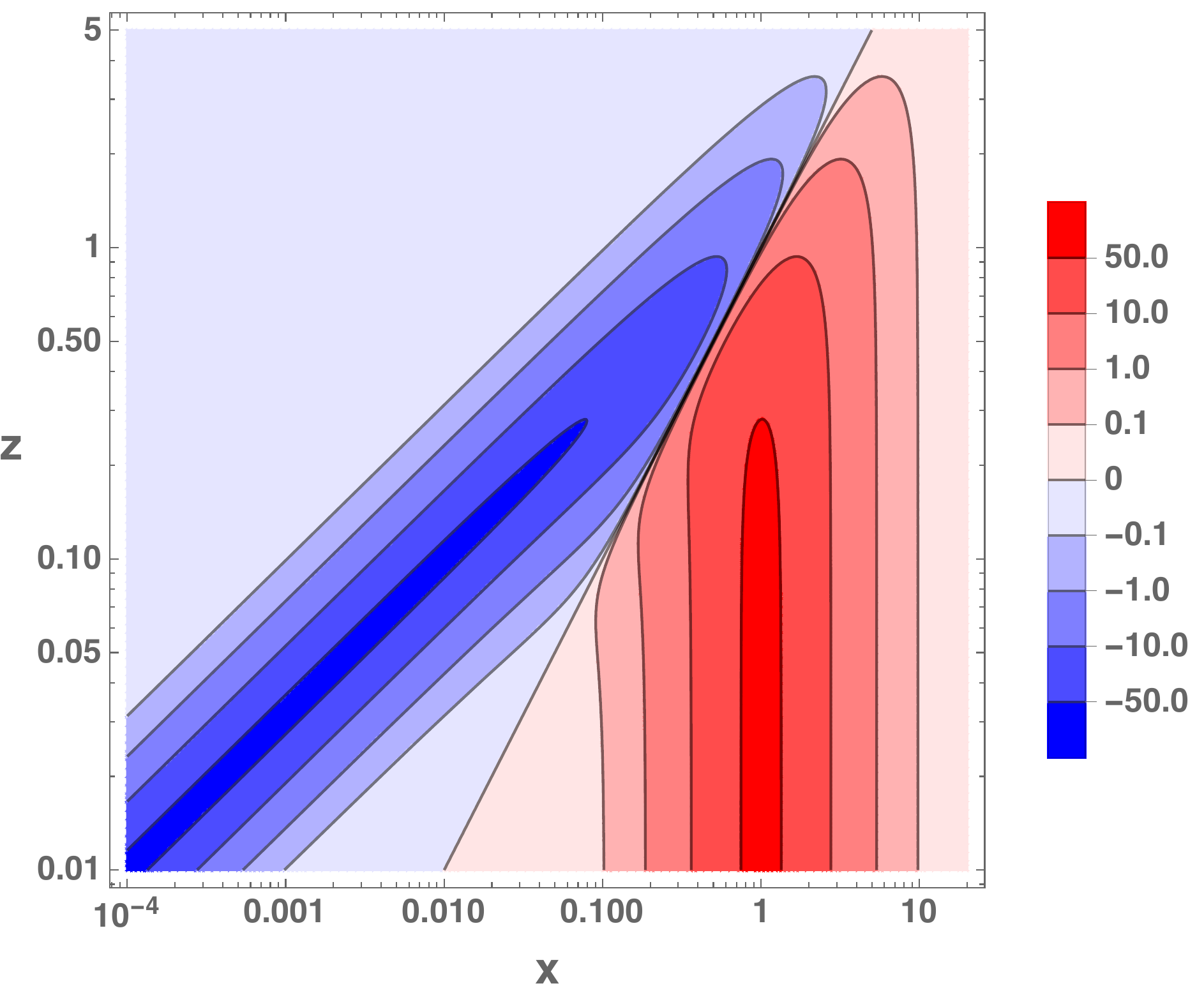}
\caption{Contour plots of the topological charge density (multiplied by the volume factor $2\pi^2 x^4$ and for $\rho=1$) of the uplifted 4D instanton (left) and the instanton-anti-instanton \eqref{eq:topcharge4DIinv} (right). In the right plot the diagonal line, $x=z$ corresponds to zero charge density with the upper (lower) region depicting contours of negative (positive) charge density associated with the anti-instanton (instanton).}
\label{fig:4dI5dinv}
\end{center}
\end{figure}

The topological charge density corresponding to \eqref{eq:4dinstantonnew2} is given by
\begin{equation}
    D = 48\frac{(x^2-z^2)(x^2+z^2)^3\rho^4}{(x^2\rho^2+(x^2+z^2)^2)^4}~.
    \label{eq:topcharge4DIinv}
\end{equation}
For any nonzero $z$, the topological charge \eqref{topologcharge} is zero (since $f(0,z)=f(\infty,z)=1$ as shown in Figure~\ref{fig:f5Dinv}), indicating that the solution consists of an instanton and anti-instanton.  Indeed, by plotting  \eqref{eq:topcharge4DIinv}  in Figure~\ref{fig:4dI5dinv} we see that, at fixed $z$, we have  an instanton of size $\rho$  and  an anti-instanton of size $z^2/\rho$ both located at $x=0$. As $z$ grows the anti-instanton size extends to $\sim \rho$ where it  pinches off (or ``annihilates") the instanton.\footnote{This exact instanton-anti-instanton solution has similarities  to the 4D solution  obtained in Ref~\cite{Lipatov:1978en,Actor:1979wv} where a complex scalar field transforming in the fundamental (doublet) representation was added to an SU(2) gauge theory, albeit with a negative quartic coupling in the scalar potential. In our case, the solution is supported by 5D  Yang-Mills fields and requires no negative couplings.}

The 5D Lagrangian density is given by
\begin{equation}
    \frac{1}{4} F_{MN}^aF^{aMN} = \frac{48\rho^4(x^2 + z^2)^4}{(x^2 \rho^2+ (x^2+z^2)^2)^4}~,
    \label{eq:energy4DIinv}
\end{equation} 
and is shown in Figure~\ref{fig:3dplot4diai}.
We can obtain the value of the action by performing the integral over $x$ first, followed by the $z$ integral, giving
\begin{equation}
        S_5= 16\pi^2 \frac{L}{g^2_5} \left(\log\frac{\rho}{\zuv} -3 \frac{\zuv^2}{\rho^2}+{\cal O}\left(\frac{\zuv^4}{\rho^4}\right)\right)~.
        \label{eq:actionxz}
\end{equation}
Again, adding the UV boundary contribution from \eqref{eq:UVLag} by substituting the solution
\eqref{eq:4dinstantonnew2},  we obtain a finite action in the limit $\zuv\to 0$:
\begin{equation}
        S_5+S_{\rm UV}= \frac{16\pi^2}{g^2_4(1/\rho)}~,
        \label{eq:finite}
\end{equation}
which consists of an instanton and anti-instanton contribution (equivalent to two instantons). However, notice that the gauge coupling is now evaluated at the scale of the instanton size $\rho$ and not at the IR scale as in \eqref{eq:5dtotal} for the uplifted 4D instanton. This follows from the fact that the solution \eqref{eq:4dinstantonnew2} is localized on the UV boundary, and extends into the bulk a distance $\sim\rho$ (see Figure~\ref{fig:f5Dinv}). This provides the expected behaviour for a Yang-Mills field interacting with matter where the running gauge coupling is evaluated at the scale $1/\rho$~\cite{Vainshtein:1981wh}.\footnote{Instead, if we perform the $z$ integration first, followed by the $x$ integration, the 5D action becomes
\begin{equation}
        S_5= 8\pi^2 \left(\log\frac{\rho}{\zuv} +\frac{1}{2} \log\frac{\rho}{x_{\rm UV}}+{\cal O}\left(\frac{z^2}{\rho^2}\right)+{\cal O}\left(\frac{x_{\rm UV}}{\rho}\right)\right)~.
         \label{eq:actionzx}
\end{equation}
There is now a logarithmic singularity as $x\rightarrow 0$. If $x_{\rm UV}=\zuv^2/\rho$ is substituted into \eqref{eq:actionzx} we recover the leading term in the action \eqref{eq:actionxz}. This is because at the scale $\zuv^2/\rho$ we encounter the anti-instanton of size $\sim \zuv^2/\rho$ located at the origin.}

The behaviour in \eqref{eq:finite} is also consistent with the supersymmetric argument given in Section~\ref{sec:uplifted4DI}. The instanton-anti-instanton configuration \eqref{eq:4dinstantonnew2} no longer preserves the accidental supersymmetry in the bulk at the classical level. This means there are no longer any fermion zero modes in the dual 4D CFT at the quantum level and therefore there is no chiral suppression in the instanton measure from the axial anomaly. Instead, above the scale $1/\rho$, there are nonzero mode contributions at the quantum level which leads to the (classical)  localized instanton-anti-instanton configuration with the running coupling behaviour in \eqref{eq:finite}.

At this point we should note that \eqref{eq:4dinstantonnew2} is only an approximate solution
of the UV-boundary equation of motion \eqref{eq:4Deombdy}  that is valid in the limit  
$1/g_4^2\gg L/g_5^2$, where the second term in \eqref{eq:4Deombdy} (CFT contribution) is neglected, and the limit $\zuv\to 0$, where \eqref{eq:4dinstantonnew2} is approximately an instanton of size $\rho$ and an anti-instanton whose size ($\zuv^2/\rho$) tends to zero:
\be
f(x,\zuv\to 0)\simeq \frac{x^2}{x^2+\rho^2}+\frac{\zuv^4/\rho^2}{x^2+\zuv^4/\rho^2}~.
\ee

Similarly, the 4D $N$-instanton solutions~\cite{tHooft:1976,Jackiw:1976fs} can also be generalized to the 5D bulk by transforming them with the 5D inversion~\eqref{eq:invtrans}. The analytic expressions are complicated but one finds anti-instantons with $z$-dependent sizes that appear at the location of the instantons, similar to the one instanton case. Thus, each instanton extends into the 5D bulk a distance $z\sim \rho_i$, where $\rho_i$ is the size of the $i$-th instanton.

\begin{figure}[t]
\begin{center}
\includegraphics[width=.30\textwidth]{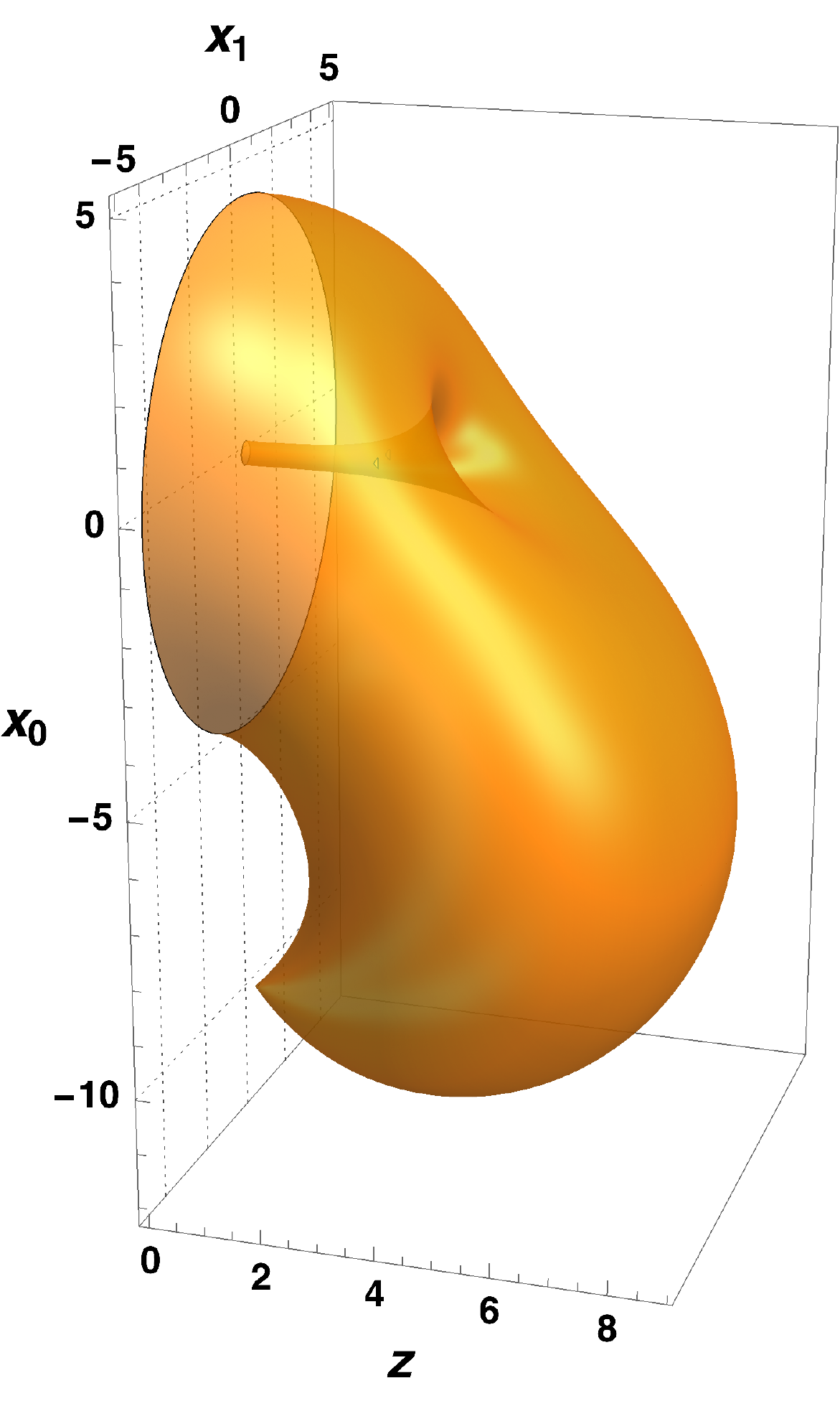}
\caption{A 2D contour of the Lagrangian density \eqref{eq:4dI_5Dsc_energyden} (multiplied by the volume factor $4\pi x^3$ and  for $\rho=1, d=10$) evaluated at 1. An anti-instanton appears at $x_0=-10$ and $x_1=x_2=x_3=0$ (Euclidean coordinates $x_{2,3}$ are not shown) whose size grows and meets
at $z\sim d$  the instanton located at $x_0=x_1=0$.}
\label{fig:4dI5dsc_energyden}
\end{center}
\end{figure}

\subsubsection{An instanton-anti-instanton by 5D special conformal transformation}

\begin{figure}[t]
\begin{center}

\includegraphics[width=.47\textwidth]{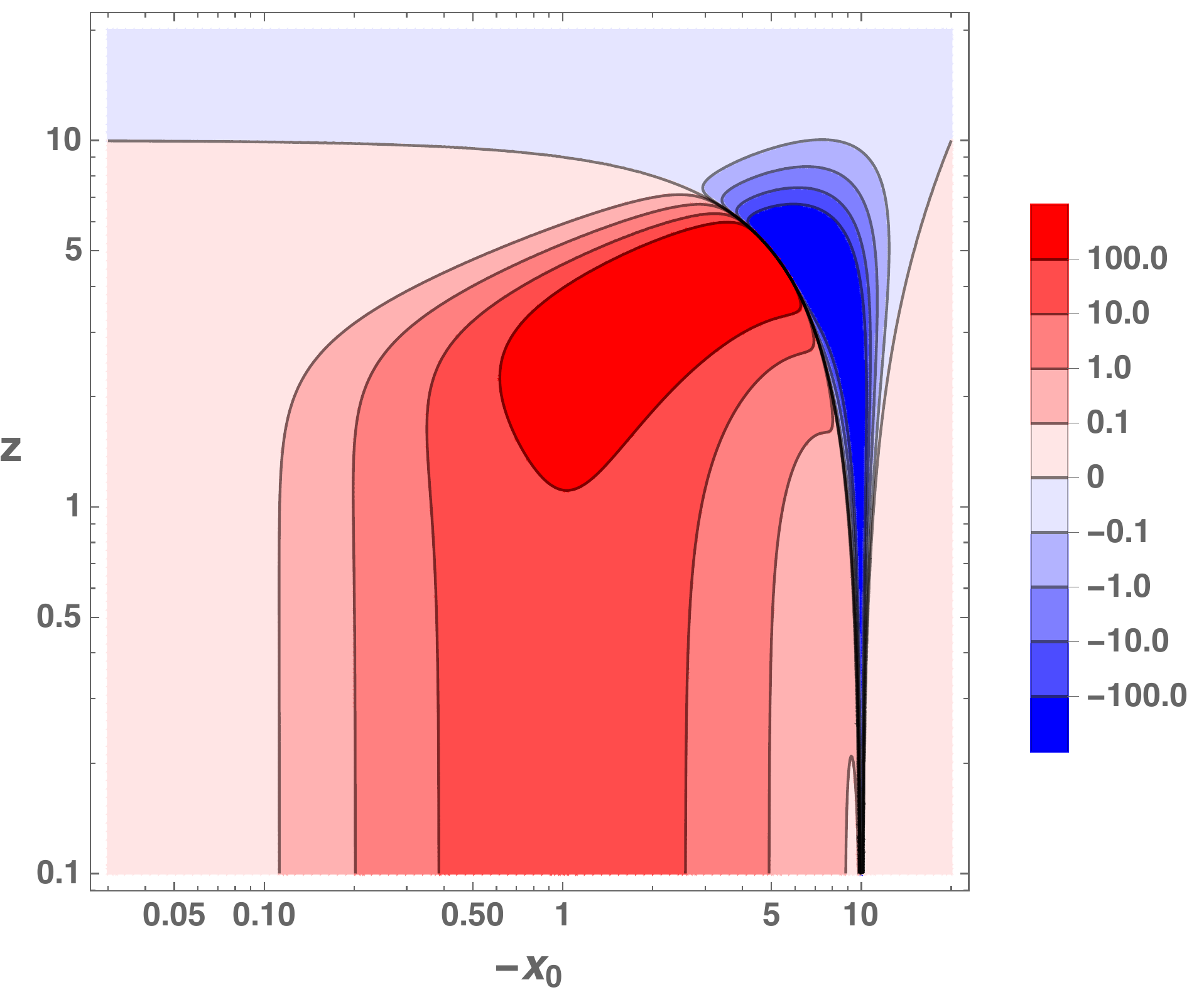}
\includegraphics[width=.52\textwidth]{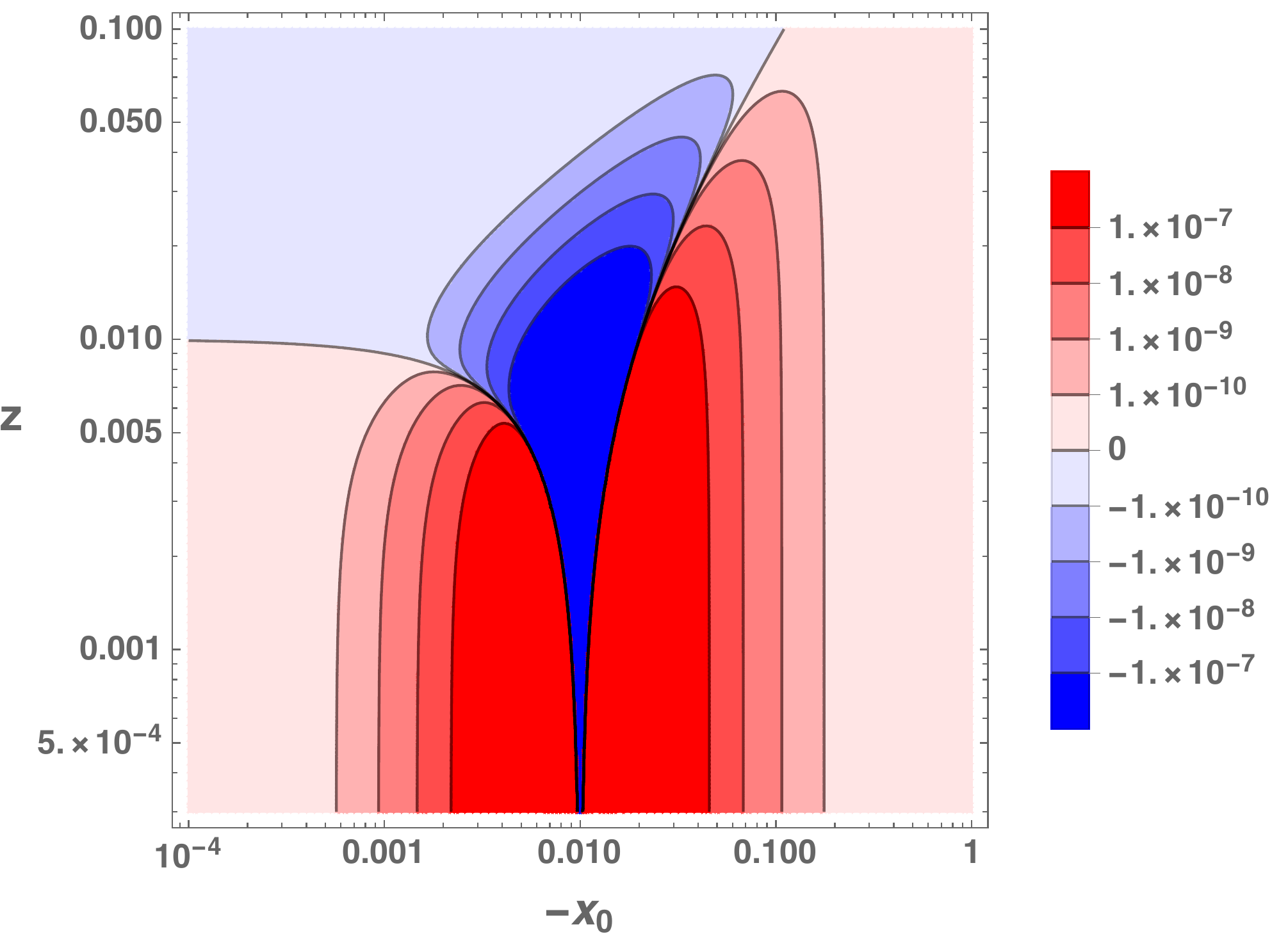}
\caption{The topological charge density \eqref{eq:4dI_5Dsc_topden} (multiplied by the volume factor 
$4\pi x^4$) of the instanton-anti-instanton \eqref{eq:sc4DI} with $\rho=1$ and $d=10\, (0.01)$ for the left (right) figure.  The contours are drawn for $x_0<0$ and $x_1=x_2=x_3=0$ and show an anti-instanton appearing at $x_0=-10\, (-0.01)$ whose size grows with increasing $z$ and annihilates the instanton located at $x_0=0$. Note the contours of the topological charge density are much smaller in the right figure.}
\label{fig:4dI5dsc}
\end{center}
\end{figure}

More general solutions can be obtained by considering special conformal transformations~\eqref{eq:conftrans}.  Without loss of generality, we can choose the parameter $d_\mu = (d,0,0,0)$ so that translations occur along the $x_0$ axis. Assuming $\delta_1=\delta_2=d$, the uplifted 4D instanton transforms to a solution dependent on $d$ and $z$ given by
\begin{equation}
A_M^a=\frac{2d^3}{h(x,z)}\begin{pmatrix} -x_1 (2 x_0+d) & r_5^2-2x_1^2 +x_0d&-2 x_1 x_2+x_3 d&  -2 x_1 x_3-x_2 d& -2 x_1 z \\
       -x_2 (2 x_0+d) & -2 x_1 x_2 -x_3 d& r_5^2-2x_2^2+x_0 d&  -2 x_2 x_3+x_1d& -2 x_2 z \\
-x_3 (2 x_0+d) &  -2 x_1 x_3  +x_2 d&-2 x_2 x_3 -x_1 d& r_5^2-2x_3^2+x_0 d& -2 x_3 z 
       \end{pmatrix}~,
       \label{eq:sc4DI}
\end{equation}
where $r_5^2=x^2+z^2$ and 
\begin{equation}
  h(x,z)=-d^4 z^2 + d^2 (x^2 + z^2)((\vec x+\vec d)^2 + z^2) + \rho^2((\vec x+\vec d)^2 + z^2 )^2 ~,
\end{equation}
with $(\vec x+\vec d)^2= x^2+2 x_0 d+ d^2$.
Note that $A_5^a$ in \eqref{eq:sc4DI} (last column) is nonzero, but when $d\rightarrow \infty$, $A_5^a\to 0$ and we recover the uplifted 4D instanton solution. The 5D Lagrangian density corresponding to \eqref{eq:sc4DI} is
\begin{equation}
\frac{1}{4} F_{MN}^aF^{aMN} = \frac{48 d^8 \rho^4 ((\vec x+\vec d)^2 + z^2 )^4}{(-d^4 z^2 + d^2 (x^2 + z^2)((\vec x+\vec d)^2 + z^2) + \rho^2((\vec x+\vec d)^2 + z^2 )^2 )^4}~.
 \label{eq:4dI_5Dsc_energyden}
\end{equation}
A two-dimensional contour of \eqref{eq:4dI_5Dsc_energyden} is shown in Figure~\ref{fig:4dI5dsc_energyden}. 
The topological charge density is
\begin{equation}
 D=\frac{48 d^8 \rho^4 ((\vec x+\vec d)^2 - z^2)((\vec x+\vec d)^2 + z^2 )^3}{(-d^4 z^2 + d^2 (x^2 + z^2)((\vec x+\vec d)^2 + z^2) + \rho^2((\vec x+\vec d)^2 + z^2 )^2 )^4}~,
       \label{eq:4dI_5Dsc_topden}
\end{equation}
and is shown in Figure~\ref{fig:4dI5dsc}. For any $z\not=0$, the topological charge is zero, indicating that 
this is an instanton-anti-instanton configuration. To understand the behaviour of the topological charge density it is useful to consider the two limits $d \gg \rho$ and $d \ll \rho$. In the first case, at any fixed value of $z\ll d$, we have an instanton of size $\rho$ located at $x=0$ with an anti-instanton of size 
$z^2/\rho$  located at $x_0=-d$ (Figure~\ref{fig:4dI5dsc}, left). In the $d \ll \rho$ limit, the instanton has a size $d^2/\rho$ and is located at $x_0 =-d$, with the anti-instanton of size $\rho z^2/d^2$ again located at $x_0=-d$ (Figure~\ref{fig:4dI5dsc}, right). In both cases, the anti-instanton grows as $z$ increases and annihilates with the instanton at $z\sim d$.

We can obtain different solutions by applying the special conformal transformation \eqref{eq:conftrans} 
using other values of $\delta_1,\delta_2$. However, they give  qualitatively similar solutions as the $\delta_1=\delta_2=d$ solution. 
Furthermore, by repeatedly applying the conformal transformation \eqref{eq:conftrans} with different constant four-vectors $d^{(\alpha)}_\mu$ $(\alpha=1,2,\dots)$, one can 
obtain multi-instanton-anti-instanton solutions.

\subsection{Meron solutions}

\subsubsection{Uplifted 4D meron}
The 5D bulk also admits meron solutions. Uplifting the well-known 4D meron~\cite{deAlfaro:1976qet,Callan:1977gz}, we have the  $z$-independent solution (for $x>0$)
\begin{equation}
      f(x,z) = \frac{1}{2}~,
      \label{eq:4dmeron}
\end{equation}
which is singular at $x=0$ with $f(x,z)=0$. This has  topological charge $q=\frac{1}{2}$. 
The Lagrangian density is
\begin{equation}
     \frac{1}{4} F_{MN}^aF^{aMN} =  \frac{3}{2x^4}~,
\end{equation}
with 5D action
\begin{equation}
      S_5=3\pi^2 \frac{L}{g_5^2} \log\left(\frac{x_{\rm IR}}{x_{\rm UV}}\right) \log\left(\frac{\zir}{\zuv}\right)~,
      \label{eq:action4Dmeron}
\end{equation}
where $x_{\rm IR\,(\rm UV)}$ are the infrared (ultraviolet) cutoffs of $x$. The 5D action is divergent in 
the $\zuv\to 0 $ limit but can be renormalized by adding the UV contribution \eqref{eq:UVLag} to give
\begin{equation}
      S_5+S_{\rm UV}=\frac{3\pi^2}{g^2_4(z_{\rm IR}^{-1})} \log\left(\frac{x_{\rm IR}}{x_{\rm UV}}\right)~.
      \label{eq:renormaction4Dmeron}
\end{equation}
This reproduces the well-known 4D meron result which has a logarithmically divergent action.
The effect of the 5D bulk is again to have the Yang-Mills coupling  evaluated at the scale $z_{\rm IR}^{-1}$ (similarly to the instanton result \eqref{eq:5dtotal}).

\begin{figure}[t]
\begin{center}
\includegraphics[width=.5\textwidth]{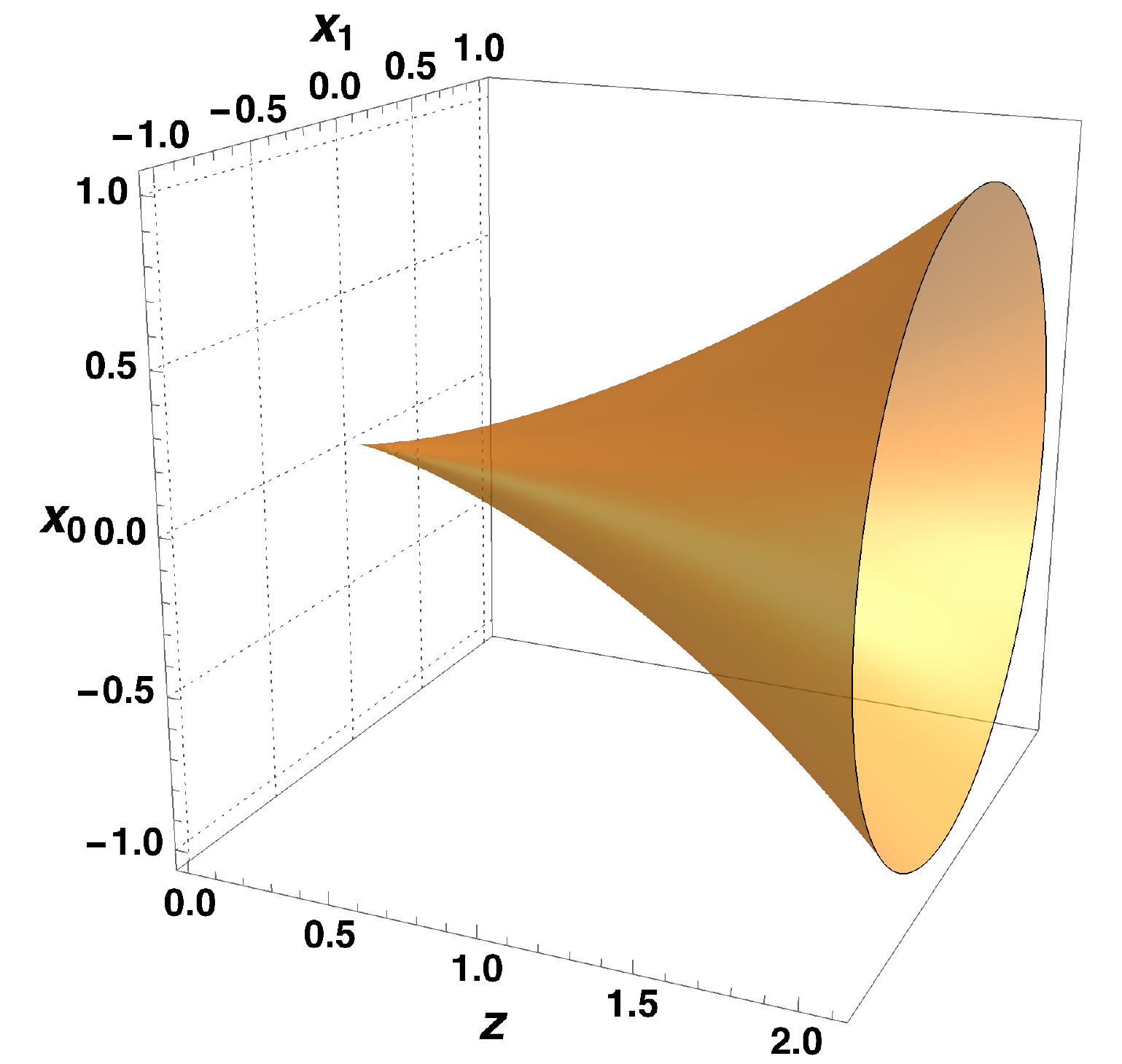}
\caption{A 2D contour of the Lagrangian density \eqref{eq:5Dm_energyden}  for the 5D meron (multiplied by the volume factor $2\pi^2 x^3$) evaluated at 10. A point-size meron appears at $x_\mu=0$  (where two Euclidean coordinates $x_{2,3}$ are not shown) which grows in size as $z$ increases.}
\label{fig:5Dm}
\end{center}
\end{figure}

\subsubsection{5D meron}
The solution \eqref{eq:4dmeron} is a trivial extension of the 4D solution since it does not depend on the 5th coordinate. It is therefore interesting to check whether nontrivial $z$-dependent solutions exist which are not obtained from a 5D conformal transformation of a known 4D solution. Remarkably, a simple and nontrivial, analytic solution of the 5D AdS Yang-Mills equation of motion \eqref{eq:feom} is found to be
\begin{equation}    
      f(x,z) = \frac{x^2}{2x^2+\frac{3}{2} z^2}~.
      \label{eq:5dmeron}
\end{equation}
For $z=0$, this solution is just the 4D meron \eqref{eq:4dmeron}  with topological charge $q=\frac{1}{2}$
that behaves as a (singular) point charge. 
However, for nonzero $z$ the topological charge density becomes nonsingular:
\begin{equation}
    D=576\frac{z^2(2 x^2+3 z^2)}{(4x^2+3z^2)^4}~.
\end{equation}
This shows that the solution \eqref{eq:5dmeron} represents a meron of finite size $\sim z$
and $q=\frac{1}{2}$, that extends into AdS$_5$. 
Similarly, an antimeron corresponds to the solution $1-f$.

\begin{figure}[t]
\begin{center}
\includegraphics[width=.5\textwidth]{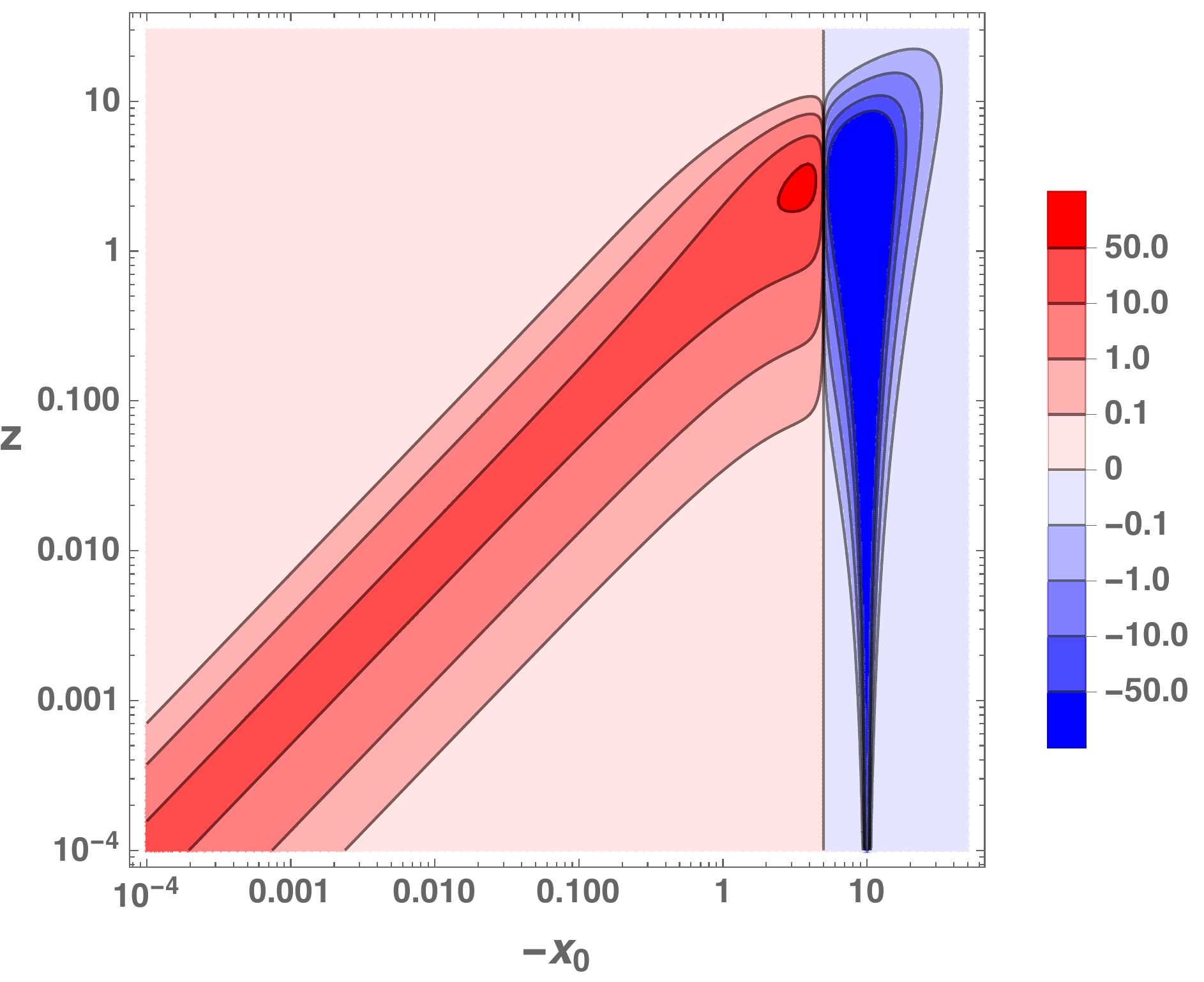}
\caption{The topological charge density \eqref{eq:5Dm_5Dsc_topden} (multiplied by the volume factor $4\pi x^4$) for  $d=10$. The contours are drawn for $x_0<0$ and $x_1=x_2=x_3=0$ and depict an antimeron appearing at $x_0=-10$ whose size grows with increasing $z$ and annihilates the meron located at $x_0=0$ whose size also grows with increasing $z$. The topological charge density is zero when $x_0=-d/2$.
}
\label{fig:5Dm_5Dsc}
\end{center}
\end{figure}

The 5D Lagrangian density is given by
\begin{equation}
    \frac{1}{4} F_{MN}^aF^{aMN} = 96 \frac{4 x^4 + 21 x^2 z^2 + 18 z^4}{(4 x^2 + 3 z^2)^4}~.
    \label{eq:5Dm_energyden}
\end{equation} 
For  $\zuv\to 0$ and $\zir \ll x_{\rm IR}$, the action becomes 
\begin{equation}
     S_5\simeq \frac{3\pi^2}{2} \frac{L}{g_5^2}\log\left(\frac{x_{\rm IR}^2}{\zir\zuv}\right)\log\left(\frac{z_{\rm IR}}{\zuv}\right)~.
\end{equation} 
Unlike the 4D meron, there is no singularity in the limit $x_{\rm UV}\rightarrow 0$, consistent with the fact that the 5D meron is a nonsingular solution for nonzero $z$.

\subsubsection{5D localized meron-antimeron solutions}

Conformal transformations can be applied to the meron solutions to generate new 5D solutions. However, the uplifted 4D meron \eqref{eq:4dmeron} and  5D meron \eqref{eq:5dmeron} are invariant under the 5D inversion and do not give new solutions. Instead, we apply the special conformation transformation to obtain $z$-dependent solutions. These solutions depend on the translation parameter $d$ that can be understood as the separation between the two single pseudoparticles.  

Consider first the 4D meron \eqref{eq:4dmeron}. Under a special conformal transformation \eqref{eq:conftrans} with $\delta_1=\delta_2=d$, the 5D Lagrangian density is
\begin{equation}
 \frac{1}{4} F_{MN}^aF^{aMN} = \frac{3 d^4}{2(d^2 z^2 - (x^2 + z^2)  ((\vec x +\vec d)^2+z^2))^2}~,
\end{equation}
while the topological charge density formally consists of a line in the bulk, connecting the meron and antimeron, that is identically zero for any given $z$.

\begin{figure}[t]
\begin{center}
\includegraphics[width=.25\textwidth]{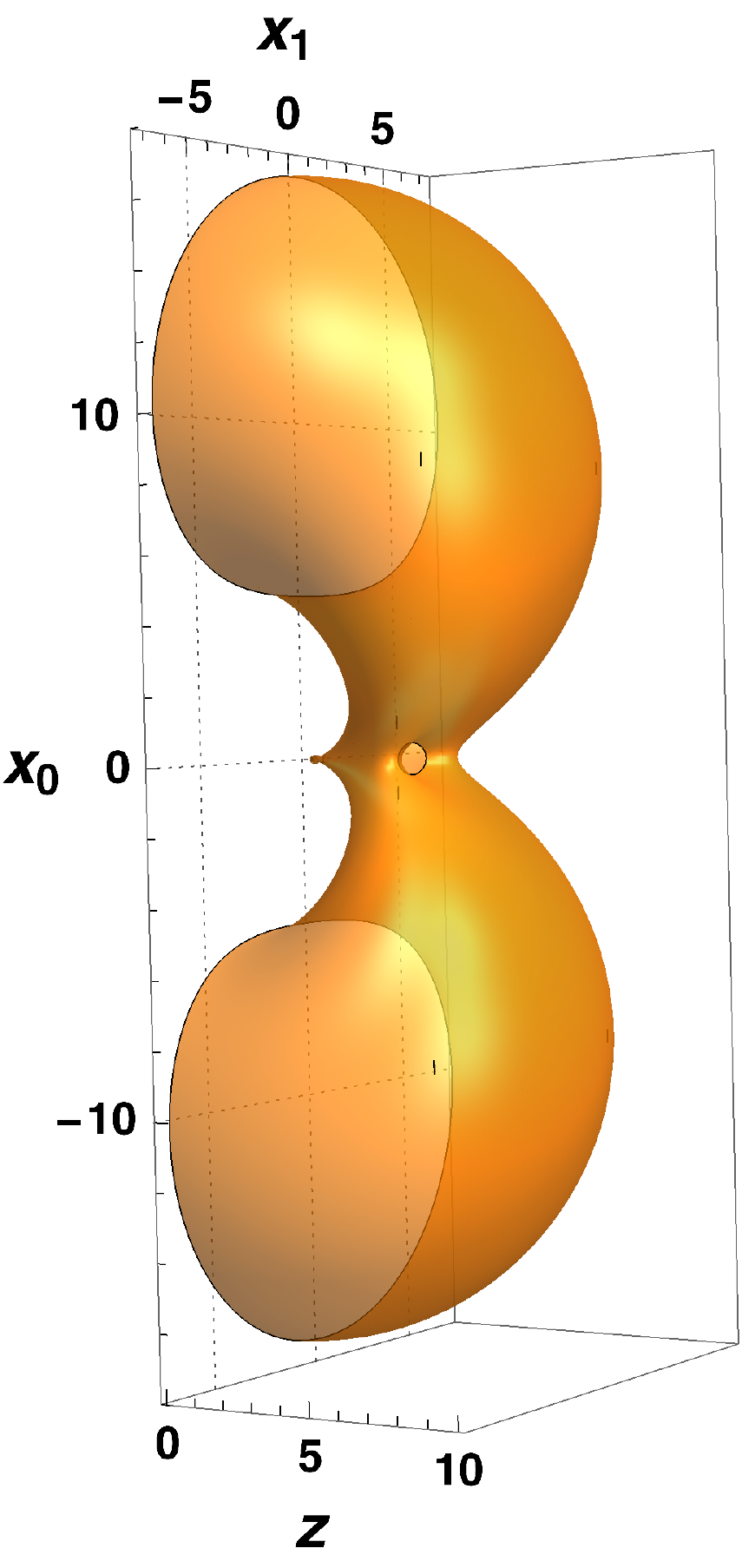}
\caption{A 2D contour of the Lagrangian density \eqref{eq:2menergy} with $b=10$ (multiplied by the volume factor $4\pi x^3$ and assuming $x_2=x_3=0$) evaluated at 10. There are two merons at $x_0=\pm 10$, and two antimerons at the origin $x_0=0$ which grow in size as $z$ increases and annihilate at $z\sim 10$.}
\label{fig:4D2m_5Dsc}
\end{center}
\end{figure}

Similarly the 5D meron \eqref{eq:5dmeron} can be transformed to a new solution by a special conformal transformation. Performing the inversions at $\delta_1=\delta_2=d$, the 5D Lagrangian density is 
\begin{equation}
\frac{1}{4} F_{MN}^aF^{aMN} = \frac{96 d^4(d^4z^4+13 d^2z^2(x^2+z^2)((\vec x+\vec d)^2+z^2)+4(x^2+z^2)^2((\vec x+\vec d)^2+z^2)^2)}{(d^2z^2-4(x^2+z^2)((\vec x+\vec d)^2+z^2))^4}~.
 \label{eq:5Dm_5Dsc_energyden}
\end{equation}
The topological density is
\begin{equation}
  D=\frac{576 d^4 z^2((\vec x+\vec d)^2-x^2)(d^2z^2+2(x^2+z^2)((\vec x+\vec d)^2+z^2))}{(d^2z^2-4(x^2+z^2)((\vec x+\vec d)^2+z^2))^4}~,
    \label{eq:5Dm_5Dsc_topden}
\end{equation}
and is shown in Figure~\ref{fig:5Dm_5Dsc}. We see that the solution corresponds to an antimeron appearing at $x_0=-d$ whose size grows with increasing $z$ and annihilates the meron located at $x_0=0$ whose size also grows with increasing $z$. When $d\rightarrow \infty$ we recover the 5D meron with $q=\frac{1}{2}$.

There are also 4D two-meron solutions that can be generalized to solutions in AdS$_5$ by transforming them under 5D conformal transformations. The 4D meron-meron solution~\cite{deAlfaro:1976qet}, 
which is just the sum of two 4D merons located at $x_0=\pm \,b$ (without loss of generality), 
can be transformed using a 5D inversion \eqref{eq:invtrans} (with $\delta=b$) to  obtain a $z$-dependent solution. Similarly, by replacing  one of the merons by an anti-meron, we can obtain a 4D meron-antimeron solution~\cite{deAlfaro:1977zs}. In both cases, after the 5D inversion, the Lagrangian density  is given by
\begin{equation}
 \frac{1}{4} F_{MN}^aF^{aMN} =\frac{24 b^4 (x^2+z^2)^4}{\left(b^4 x^4+(x^2+z^2)^4-2 b^2 (2x_0^2-x^2)(x^2+z^2)^2  \right)^2}~.
     \label{eq:2menergy}
\end{equation}
A 2D contour of \eqref{eq:2menergy} is shown in Figure~\ref{fig:4D2m_5Dsc} for $b=10$, which depicts two merons at $x_0=\pm 10$ and two antimerons at the origin $x_0=0$ which grow in size as $z$ increases, and annihilate at $z\sim 10$.

A special conformal transformation can also be performed on both 4D meron-meron and meron-antimeron solutions with independent translation parameters $d_\mu^{(\alpha)} (\alpha=1,2,\dots)$, leading to multi-meron-anti-meron solutions similar to the instanton case.

\section{Conclusion}

Weakly-gauged holographic models, which are dual to Yang-Mills theories interacting with a strongly-coupled CFT, are an interesting class of theories in which to study the effects of small instantons, since the gauge coupling of the UV-localized gluon can grow at small distances without introducing extra colored matter. However, contrary to expectations, we show that this feature is nullified by topological charge conservation which forbids UV localized instanton configurations and causes small instanton effects to be highly suppressed. Surprisingly, given that the 5D theory has, in principle, no knowledge of the underlying constituents in the dual 4D theory, we find that this suppression is equivalent to that obtained from a 4D dual theory with colored fermion constituents. 

In fact this result can be understood by noting that the ``cylinder" instanton configuration remains a solution in a supersymmetric version of the 5D model, leading to an {\it accidental} supersymmetry in the bulk. By the AdS/CFT dictionary, the corresponding dual 4D theory is then a supersymmetric CFT that necessarily contains colored fermion constituents. At the quantum level, the supersymmetric CFT contributes fermion zero modes to the gauge coupling running while the nonzero mode contributions cancel. Thus the chiral suppression from the fermion zero modes in the instanton measure can be neatly interpreted as a running in the gauge coupling from $1/\rho$ to $\zir^{-1}$ to give the ``cylinder" instanton action $S_E=8\pi^2/g^2(\zir^{-1})$ and not the smaller action $S_E=8\pi^2/g^2(1/\rho)$ from a UV-localized instanton.

Despite the absence of a UV localized instanton solution, we discover other new solutions of the AdS$_5$ Yang-Mills equations that are UV-localized but have zero topological charge. These solutions are simply obtained by performing 5D conformal transformations on known 4D solutions. This includes an instanton-anti-instanton solution (obtained from a 5D inversion on the 4D instanton) where an anti-instanton of size $z^2/\rho$ grows inside an instanton of size $\rho$ until they annihilate at a distance $z\sim \rho$ in the 5D bulk. Other instanton-anti-instanton solutions can be obtained by performing 5D special conformal transformations on the 4D instanton and feature non-spherically symmetric configurations. In fact the solution generating technique can be extended to obtain an arbitrary number of instanton-anti-instanton configurations.

These localized zero topological charge solutions can also be interpreted from the perspective of  accidental supersymmetry in the bulk. The instanton-anti-instanton configurations do not preserve the 5D supersymmetry, and therefore the corresponding dual 4D CFT does not contain any fermion zero modes 
at the quantum level. This means that the contribution to the running gauge coupling is only from nonzero modes above the scale $1/\rho$, which is consistent with the localized 5D solutions that are truncated at a distance $z\sim \rho$ in the bulk.

Moreover, we can apply 5D conformal transformations on known 4D meron and multi-meron solutions. This again gives rise to UV localized meron-antimeron configurations with zero topological charge.
However, we also discovered a new 5D meron solution that is not obtained from any 5D conformal transformation on a known 4D solution. At any fixed $z$ in the AdS bulk, it has topological 
charge $1/2$ and is regular. Furthermore, performing 5D conformal transformations on the 5D meron solution leads to new meron-antimeron configurations.

Our analysis has made a number of simplifying assumptions on the 5D AdS Yang-Mills theory. We have only studied smooth solutions in the bulk, neglected gravitational interactions and worked in the limit $g_5\rightarrow 0$ such that the 5D cutoff $\Lambda_5 \rightarrow \infty$. Also at the UV boundary we have neglected bulk effects. We expect that including these effects will modify our solutions although the qualitative behaviour should remain unchanged. 

Despite these simplifications, our work has uncovered new nonperturbative configurations of Yang-Mills theories interacting with strongly-coupled CFTs. These solutions can be understood from a bulk topological charge conservation or an accidental supersymmetry in the bulk that relates the presence of UV localized (classical) 5D configurations to the quantum corrections in the dual 4D theory.
While accidental supersymmetry necessarily implies colored fermion constituents in the dual 4D theory, it remains unclear whether nonsupersymmetric CFTs with colored scalar zero modes could also exist to explain our results. In any case, the holographic approach has provided some answers and opened a new way to study Euclidean Yang-Mills solutions.

\section*{Acknowledgments}
We thank Aleksey Cherman, Erich Poppitz, Misha Shifman and Arkady Vainshtein for helpful discussions. The work of T.G. is supported in part by the DOE Grant No. DE-SC0011842 at the University of Minnesota, and the Simons Foundation. The work of A.P. is supported by the Catalan ICREA Academia Program and  grants FPA2017-88915-P, 2017-SGR-1069 and SEV-2016-0588.  T.G. acknowledges the Aspen Center for Physics which is supported by the National Science Foundation grant PHY-1607611, where part of this work was done.

\appendix
\section{5D Conformal Transformations}
\label{sec:App5D}

Consider the transformation $ x_M \rightarrow  {\bar x}_M(x,z)$ where the metric \eqref{eq:5dmetric} is invariant i.e. $g_{MN}(x) =g_{MN}(\bar x)$. Then the action satisfies
\begin{equation}
   \int d^5 {\bar x} \sqrt{g({\bar x})}~\frac{1}{4 g_5^2} F_{MN}^a({\bar x}) F^{aMN}({\bar x})= \int d^5 x \sqrt{g(x)}~\frac{1}{4 g_5^2} F_{MN}^a(x) F^{aMN}(x)~,
   \label{eq:5Dinv}
\end{equation}
where 
\begin{eqnarray}
        A_M^a(x) &=& \frac{\partial {\bar x}^N}{\partial x_M} A_N^a({\bar x})~,\label{eq:Atransform}\\
        F_{MN}^a(x)&=& \frac{\partial {\bar x}^R}{\partial x_M}\frac{\partial {\bar x}^S}{\partial x_N} 
        F_{RS}^a({\bar x})~.     
\end{eqnarray}
Thus if $A_M^a({\bar x})$ is a solution to the 5D Yang-Mills equations of motion \eqref{eq:5Deom}, then given \eqref{eq:5Dinv}, $A_M^a(x)$ will also be a solution.\footnote{Note that the new solution gives a different value of the action because the integration limits in \eqref{eq:5Dinv} are not transformed. Also the new solution can have a different topological charge, since the topological charge density is not invariant under the 5D conformal transformation.}  Note that if $\frac{\partial {\bar x}^N}{\partial x_M} =\delta_{MN}$ then no new solution is generated. 

We next summarize the possible isometries of AdS$_5$. These consist of discrete and continuous transformations which keeps the metric invariant.

\subsection{Discrete Transformations}
A discrete isometry of AdS$_5$ space is the 5D coordinate inversion, given by
\begin{equation}
        x_M \rightarrow {\bar x}_M(x,z) =  \frac{\delta^2} {x^2+z^2}x_M~,
        \label{eq:invtrans}
\end{equation}
where the inversion is performed at the distance scale (or radius) $\delta$. A useful relation is
\begin{equation}
 {\bar r}_5^2\equiv {\bar x}_M {\bar x}^M = \frac{\delta^4}{x^2+z^2}~.
\end{equation}

\subsection{Continuous Transformations}
The continuous isometry group of Euclidean AdS$_5$ is $SO(5,1)$ which is a 15 parameter group consisting of the 10 parameter Poincare group (of 4D translations, 3D rotations and Lorentz boosts) and an additional 5 parameters that parameterize dilations and special conformal tranformations. These are given by

\subsubsection{Dilation}
\label{sec:dilation}
This consists of the transformation
\begin{equation}
        x_M \rightarrow {\bar x}_M(x,z) = \lambda \,x_M~,
        \label{eq:conftransdil}
\end{equation}
where $\lambda$ is a real parameter. Note that two successive discrete inversions \eqref{eq:invtrans}, first with radius $\delta_1$, followed by a second with radius $\delta_2$, is equivalent to a dilation with $\lambda=\frac{\delta_2^2}{\delta_1^2}$.

\subsubsection{Special conformal transformation}
This transformation is given by
\begin{equation}
        x_M \rightarrow {\bar x}_M(x,z) = \frac{\delta_2^2}{\delta_1^2} \frac{x_M+ d_M\frac{1}{\delta_1^2} (x^2+z^2)}{\left(1+ 2 \frac{d\cdot x}{\delta_1^2} + \frac{d^2}{\delta_1^4} (x^2+z^2)\right)}~,
        \label{eq:conftrans}
\end{equation}
where $d_M=(d_\mu,0)$ is an arbitrary constant vector in the $x_\mu$ direction. The special conformal transformation corresponds to an inversion at the distance scale $\delta_1$, followed by a translation ($x_M\to x_M+d_M$), then a final inversion at the distance scale, $\delta_2$. The radial Euclidean distance-squared transforms as 
\begin{equation}
 {\bar r}_5^2\equiv {\bar x}_M {\bar x}^M = \frac{\delta_2^4}{\delta_1^4}\frac{x^2+z^2}{\left(1+ 2 \frac{d\cdot x}{\delta_1^2} + \frac{d^2}{\delta_1^4} (x^2+z^2)\right)}~.
\end{equation}
Note that the metric \eqref{eq:5dmetric} remains invariant under \eqref{eq:conftrans} while the gauge fields transform as \eqref{eq:Atransform}.

\bibliographystyle{JHEP}
\bibliography{references}

\providecommand{\href}[2]{#2}\begingroup\raggedright\begin{thebibliography}{10}

\bibitem{Belavin:1975fg}
A.~A. Belavin, A.~M. Polyakov, A.~S. Schwartz and {\relax Yu}.~S. Tyupkin,
  \emph{{Pseudoparticle Solutions of the Yang-Mills Equations}},
  \href{https://doi.org/10.1016/0370-2693(75)90163-X}{\emph{Phys. Lett.}
  {\bfseries 59B} (1975) 85}.

\bibitem{Agrawal:2017ksf}
P.~Agrawal and K.~Howe, \emph{{Factoring the Strong CP Problem}},
  \href{https://doi.org/10.1007/JHEP12(2018)029}{\emph{JHEP} {\bfseries 12}
  (2018) 029} [\href{https://arxiv.org/abs/1710.04213}{{\ttfamily
  1710.04213}}].

\bibitem{Agrawal:2017evu}
P.~Agrawal and K.~Howe, \emph{{A Flavorful Factoring of the Strong CP
  Problem}}, \href{https://doi.org/10.1007/JHEP12(2018)035}{\emph{JHEP}
  {\bfseries 12} (2018) 035}
  [\href{https://arxiv.org/abs/1712.05803}{{\ttfamily 1712.05803}}].

\bibitem{Csaki:2019vte}
C.~Cs\'aki, M.~Ruhdorfer and Y.~Shirman, \emph{{UV Sensitivity of the Axion
  Mass from Instantons in Partially Broken Gauge Groups}},
  \href{https://doi.org/10.1007/JHEP04(2020)031}{\emph{JHEP} {\bfseries 04}
  (2020) 031} [\href{https://arxiv.org/abs/1912.02197}{{\ttfamily
  1912.02197}}].

\bibitem{Gherghetta:2020keg}
T.~Gherghetta, V.~V. Khoze, A.~Pomarol and Y.~Shirman, \emph{{The Axion Mass
  from 5D Small Instantons}},
  \href{https://doi.org/10.1007/JHEP03(2020)063}{\emph{JHEP} {\bfseries 03}
  (2020) 063} [\href{https://arxiv.org/abs/2001.05610}{{\ttfamily
  2001.05610}}].

\bibitem{Pomarol:2000hp}
A.~Pomarol, \emph{{Grand unified theories without the desert}},
  \href{https://doi.org/10.1103/PhysRevLett.85.4004}{\emph{Phys. Rev. Lett.}
  {\bfseries 85} (2000) 4004}
  [\href{https://arxiv.org/abs/hep-ph/0005293}{{\ttfamily hep-ph/0005293}}].

\bibitem{Arkani-Hamed:2000ijo}
N.~Arkani-Hamed, M.~Porrati and L.~Randall, \emph{{Holography and
  phenomenology}},
  \href{https://doi.org/10.1088/1126-6708/2001/08/017}{\emph{JHEP} {\bfseries
  08} (2001) 017} [\href{https://arxiv.org/abs/hep-th/0012148}{{\ttfamily
  hep-th/0012148}}].

\bibitem{Goldberger:2002cz}
W.~D. Goldberger and I.~Z. Rothstein, \emph{{High-energy field theory in
  truncated AdS backgrounds}},
  \href{https://doi.org/10.1103/PhysRevLett.89.131601}{\emph{Phys. Rev. Lett.}
  {\bfseries 89} (2002) 131601}
  [\href{https://arxiv.org/abs/hep-th/0204160}{{\ttfamily hep-th/0204160}}].

\bibitem{Maldacena:1997re}
J.~M. Maldacena, \emph{{The Large N limit of superconformal field theories and
  supergravity}}, \href{https://doi.org/10.1023/A:1026654312961}{\emph{Adv.
  Theor. Math. Phys.} {\bfseries 2} (1998) 231}
  [\href{https://arxiv.org/abs/hep-th/9711200}{{\ttfamily hep-th/9711200}}].

\bibitem{Actor:1979in}
A.~Actor, \emph{{Classical Solutions of SU(2) Yang-Mills Theories}},
  \href{https://doi.org/10.1103/RevModPhys.51.461}{\emph{Rev. Mod. Phys.}
  {\bfseries 51} (1979) 461}.

\bibitem{deAlfaro:1976qet}
V.~de~Alfaro, S.~Fubini and G.~Furlan, \emph{{A New Classical Solution of the
  Yang-Mills Field Equations}},
  \href{https://doi.org/10.1016/0370-2693(76)90022-8}{\emph{Phys. Lett. B}
  {\bfseries 65} (1976) 163}.

\bibitem{Callan:1977gz}
C.~G. Callan, Jr., R.~F. Dashen and D.~J. Gross, \emph{{Toward a Theory of the
  Strong Interactions}},
  \href{https://doi.org/10.1103/PhysRevD.17.2717}{\emph{Phys. Rev.} {\bfseries
  D17} (1978) 2717}.

\bibitem{deAlfaro:1977zs}
V.~de~Alfaro, S.~Fubini and G.~Furlan, \emph{{Properties of O(4) x O(2)
  Symmetric Solutions of the Yang-Mills Field Equations}},
  \href{https://doi.org/10.1016/0370-2693(77)90703-1}{\emph{Phys. Lett. B}
  {\bfseries 72} (1977) 203}.

\bibitem{tHooft:1976snw}
G.~'t~Hooft, \emph{{Computation of the Quantum Effects Due to a
  Four-Dimensional Pseudoparticle}},
  \href{https://doi.org/10.1103/PhysRevD.14.3432}{\emph{Phys. Rev. D}
  {\bfseries 14} (1976) 3432}.

\bibitem{Vainshtein:1981wh}
A.~I. Vainshtein, V.~I. Zakharov, V.~A. Novikov and M.~A. Shifman, \emph{{ABC's
  of Instantons}},
  \href{https://doi.org/10.1070/PU1982v025n04ABEH004533}{\emph{Sov. Phys. Usp.}
  {\bfseries 25} (1982) 195}.

\bibitem{Gherghetta:2000qt}
T.~Gherghetta and A.~Pomarol, \emph{{Bulk fields and supersymmetry in a slice
  of AdS}}, \href{https://doi.org/10.1016/S0550-3213(00)00392-8}{\emph{Nucl.
  Phys. B} {\bfseries 586} (2000) 141}
  [\href{https://arxiv.org/abs/hep-ph/0003129}{{\ttfamily hep-ph/0003129}}].

\bibitem{Shifman:1979uw}
M.~A. Shifman, A.~I. Vainshtein and V.~I. Zakharov, \emph{{Instanton Density in
  a Theory with Massless Quarks}},
  \href{https://doi.org/10.1016/0550-3213(80)90389-2}{\emph{Nucl. Phys. B}
  {\bfseries 163} (1980) 46}.

\bibitem{Shifman:1999mv}
M.~A. Shifman and A.~I. Vainshtein, \emph{{Instantons versus supersymmetry:
  Fifteen years later}},  pp.~485--647, 2, 1999,
  \href{https://arxiv.org/abs/hep-th/9902018}{{\ttfamily hep-th/9902018}}.

\bibitem{Poppitz:2002ac}
E.~Poppitz and Y.~Shirman, \emph{{The Strength of small instanton amplitudes in
  gauge theories with compact extra dimensions}},
  \href{https://doi.org/10.1088/1126-6708/2002/07/041}{\emph{JHEP} {\bfseries
  07} (2002) 041} [\href{https://arxiv.org/abs/hep-th/0204075}{{\ttfamily
  hep-th/0204075}}].

\bibitem{Lipatov:1978en}
L.~N. Lipatov, A.~P. Bukhvostov and E.~I. Malkov, \emph{{Large Order Estimates
  for Perturbation Theory of a Yang-Mills Field Coupled to a Scalar Field}},
  \href{https://doi.org/10.1103/PhysRevD.19.2974}{\emph{Phys. Rev. D}
  {\bfseries 19} (1979) 2974}.

\bibitem{Actor:1979wv}
A.~Actor, \emph{{Properties of an Instanton Anti-instanton Solution of a
  {Yang-Mills} Theory}}, \href{https://doi.org/10.1007/BF01414188}{\emph{Z.
  Phys. C} {\bfseries 3} (1979) 353}.

\bibitem{tHooft:1976}
G.~'t~Hooft, \emph{unpublished}, .

\bibitem{Jackiw:1976fs}
R.~Jackiw, C.~Nohl and C.~Rebbi, \emph{{Conformal Properties of Pseudoparticle
  Configurations}}, \href{https://doi.org/10.1103/PhysRevD.15.1642}{\emph{Phys.
  Rev. D} {\bfseries 15} (1977) 1642}.

\end{thebibliography}\endgroup

\end{document}